\documentclass[conference]{IEEEtran}

\IEEEoverridecommandlockouts

\usepackage{cite}
\usepackage{amsmath,amssymb,amsfonts}
\usepackage{algorithmic}
\usepackage{graphicx}
\usepackage{textcomp}
\usepackage{xcolor}

\usepackage{url}
\IEEEtriggeratref{5}

\def\BibTeX{{\rm B\kern-.05em{\sc i\kern-.025em b}\kern-.08em
    T\kern-.1667em\lower.7ex\hbox{E}\kern-.125emX}}

\begin{document}

\title{Niel's Chess: A Quantum Game for Schools and the General Public}

\author{\IEEEauthorblockN{Tam\'as Varga}
\IEEEauthorblockA{\textit{Constructor Institute Schaffhausen}\\
Schaffhausen, Switzerland \\
tamas.varga@constructor.org}
}

\maketitle

\begin{abstract}
In this paper, a quantum variant of chess is introduced, which can be played on a traditional board, without using computers or other electronic devices. The rules of the game arise naturally by combining the rules of conventional chess with key quantum-physical effects such as superposition and entanglement. Niel's Chess is recommended for ages 10 and above, to everyone who wishes to play a creative game with historical roots and at the same time gain intuition about the foundational quantum effects that power cutting-edge technologies like quantum computing and quantum communication, which are poised to revolutionize our society in the coming decades. Takeaways from a pilot educational session that was carried out with 10-to-12-year-old children are also presented.
\end{abstract}


\section{Introduction}\label{sec1}

Quantum information technology is on the cusp of disrupting a wide range of industries~\cite{mckinsey}. In the next decades, quantum computers are expected to help solve certain problems of great importance that would otherwise be infeasible using conventional computers only. Moreover, quantum-physical effects make it possible to devise communication protocols whose security is guaranteed by the laws of physics, as opposed to relying on mathematical problems that are believed, but not proved, to be hard to solve.

At the heart of the upcoming quantum revolution lies the quantum effect \textit{superposition}, which means that a physical system is in a state which is a combination of two or more states that are mutually exclusive. For example, according to quantum theory, an atom's location can be a combination of two or more different locations, somehow as if the atom was in multiple places at the same time. However, when someone tries to observe the atom to see what it looks like being at two or more locations at once, the superposition immediately 'collapses' and the atom will be found at exactly one of those locations, picked by nature in a truly random fashion; nobody in the universe can predict the outcome with certainty. What is more, superposition leads to the even stranger quantum effect \textit{entanglement}, where two or more distinct, typically spatially separated physical systems seem to be connected, despite having no apparent physical link between them. It is as if two coins tossed in two far-away cities were destined to (randomly) land on the same side.

As quantum information technologies become more and more widespread in the future, various quantum concepts will gradually penetrate into our professional as well as personal lives. Niel's Chess can be an example of the latter, a traditional board game in which the key to success is an informal acquaintance with superposition and entanglement. Besides being an interesting logic game, Niel's Chess can also serve educational purposes, helping educators prepare the wide public, especially the younger generations, for the quantum era our society is heading towards.

While many games have been considered as a means to facilitate quantum technology education and outreach~\cite{seskir22,piispanen23,artner23,kopf23,qodyssey,nita21,quantime22,qplaylearn,uchicago,scienceathome}, chess stands out due to its worldwide popularity, both as a game and a sport, and its established place in human culture. Thus, a quantum game which is derived from chess has the potential that people will continue playing it outside the classroom, simply because it is fun, feels familiar and poses a (no-math) challenge to the brain.

\section{Previous Works}\label{sec2}

Previous attempts to extend chess to the quantum realm include, most notably, Akl~\cite{akl16} and Cantwell~\cite{cantwell19}.

In~\cite{akl16}, the idea is that each piece on the board, except the king, can be in a superposition state, being of two types at once. For example, a piece may be a knight and a bishop simultaneously. Now, if the player touches the piece to make a move, it 'collapses' with equal probability, becoming either a knight or a bishop, and then the move can be made with it accordingly. In case there is no possible move after the collapse, the player's turn is over. Furthermore, if a piece lands on a black square, it regains its original superposition state, otherwise it remains in a conventional (collapsed) state. Every piece except the king starts out in a random superposition state, that is, being of two types at once, assigned by the computer based on a random scheme. Initially, neither of the players know in what superposition any given piece exactly is; they gain information about that only by seeing the piece collapse. In the most extreme case, a piece might be in a superposition of being a pawn and a queen. The king may be placed, or left, in check, and the game ends when a player captures the opponent's king. Subjectively, all these rules seem to represent a substantial departure from the mindset of conventional chess.

In~\cite{cantwell19}, a different philosophy is used. Instead of the individual pieces, it is the whole game that is in a superposition state. Namely, it is a superposition of conventional chessboards, each having a different position of the pieces on it, as if multiple related chess games were being played in parallel. A computer keeps track of the superposition state of the game, and there is a specific (partial) collapse rule, executed by the system whenever necessary before a move is made, which uses a scheme to (randomly) remove boards from the superposition, so as to prevent the overall situation from becoming unmanageable in terms of execution and visual representation. There is no notion of check or checkmate, and similarly to~\cite{akl16} the king can be captured like any other piece. The game ends as soon as a player does not have a king on any of the boards in the superposition. One difficulty for the player is that in some cases it is necessary to know the exact mathematical calculations, which the computer performs in the background using complex numbers, to understand why certain positions arise.

Both of the aforementioned quantum variants of chess are computer games, and so are the majority of quantum games used in quantum technology education. In contrast, Niel's Chess is a traditional board game that does not need electricity, let alone computers, allowing people to play the game wherever conventional chess can be played, and also giving them more responsibility from the outset in applying the rules correctly. Moreover, the chess set is easy to make in a DIY\footnote{DIY = Do it yourself} manner (see Appendix~\ref{secA}) and there is no need to look at a digital screen while playing.

\section{Niel's Chess}\label{sec3}

In this section, the main ideas and rules of Niel's Chess\texttrademark\ are explained, which served as the basis of a pilot educational session, described in Section~\ref{sec4}, in which a subset of those ideas and rules were taught to a small group of children aged 10 to 12. For a complete specification of the rules (including details such as 'en passant' and castling), adhering to the format in the FIDE\footnote{FIDE = International Chess Federation} Laws of Chess document \cite{fide22}, the reader is referred to~\cite{varga23intl}.

\subsection{Design Principles}\label{subsec31}

The quantum-chess variant presented here takes a different approach compared to~\cite{akl16} and~\cite{cantwell19}. Below is a list of the most important guiding principles, while Appendix~\ref{secB} discusses technical differences in addition to these principles:

\begin{itemize}
\item \textbf{Traditional: }the game should be played without relying on computers or other electronic devices.
\item \textbf{It is chess: }preserve the characteristics of chess pieces, their movements, the notion of check, and the visibility of the game's state for both players.
\item \textbf{Logic game: }the quantum world is inherently random, but luck should not make up for lack of skill.
\item \textbf{Intuitive: }make quantum rules arise naturally by combining quantum effects with the rules of chess.
\end{itemize}

As a result, no quantum effect is introduced just for the sake of it. All quantum rules are simple and harmonize with the characteristics and spirit of chess. One crucial example is that the king cannot be captured, only checkmated. Furthermore, the complete state of the game can be seen by looking at the chessboard. Nothing is hidden, or difficult to calculate. The addition of quantum rules enriches chess with creative ideas in both attack and defense.

To lower the barrier to entry for newcomers, in particular children, it is recommended to start by playing the game on smaller boards (see Appendix~\ref{secC}) and with a reduced set of rules.

\subsection{Indefinite Pieces}\label{subsec32}

\begin{figure}[t]
\centering
\includegraphics[width=0.48\columnwidth]{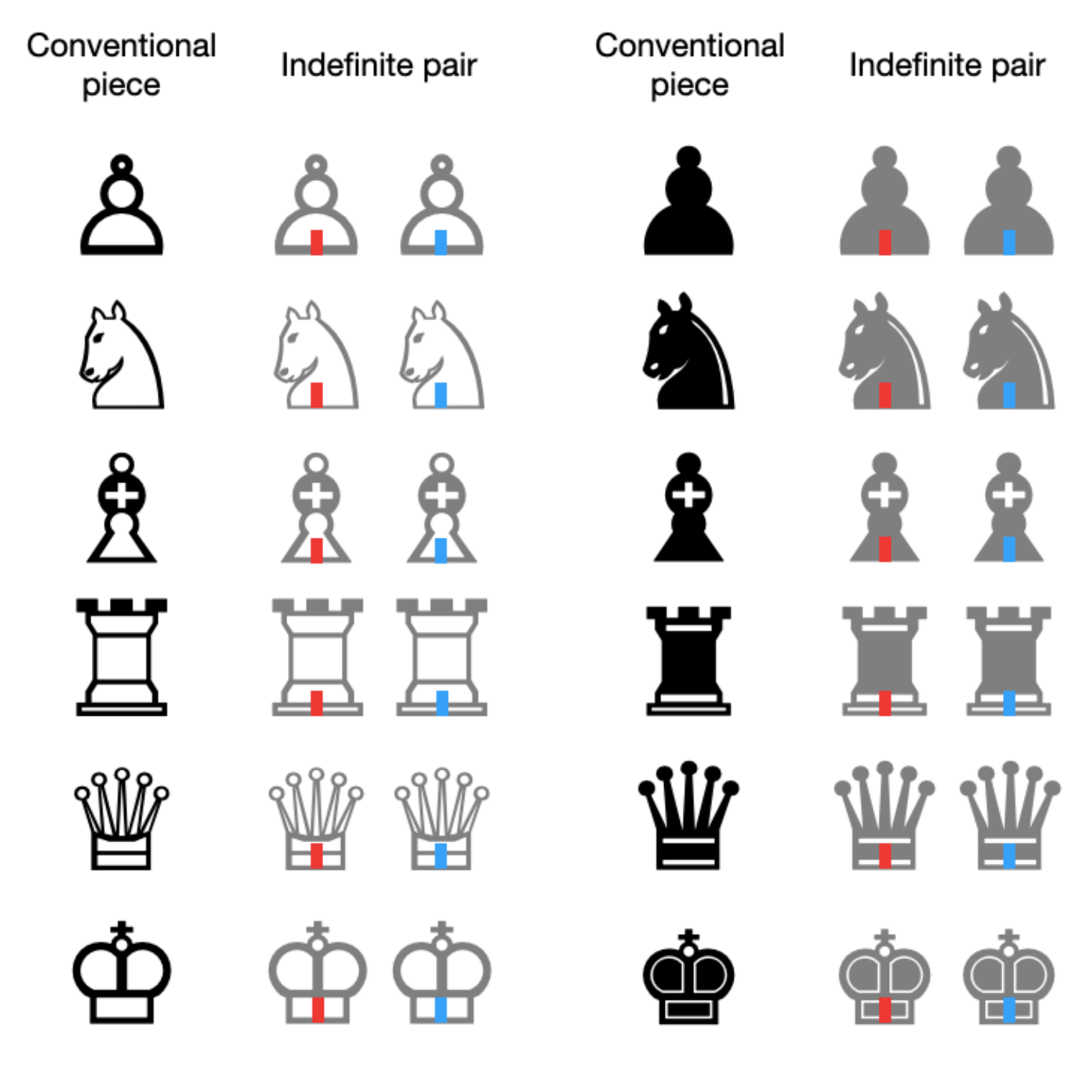}\quad\includegraphics[width=0.48\columnwidth]{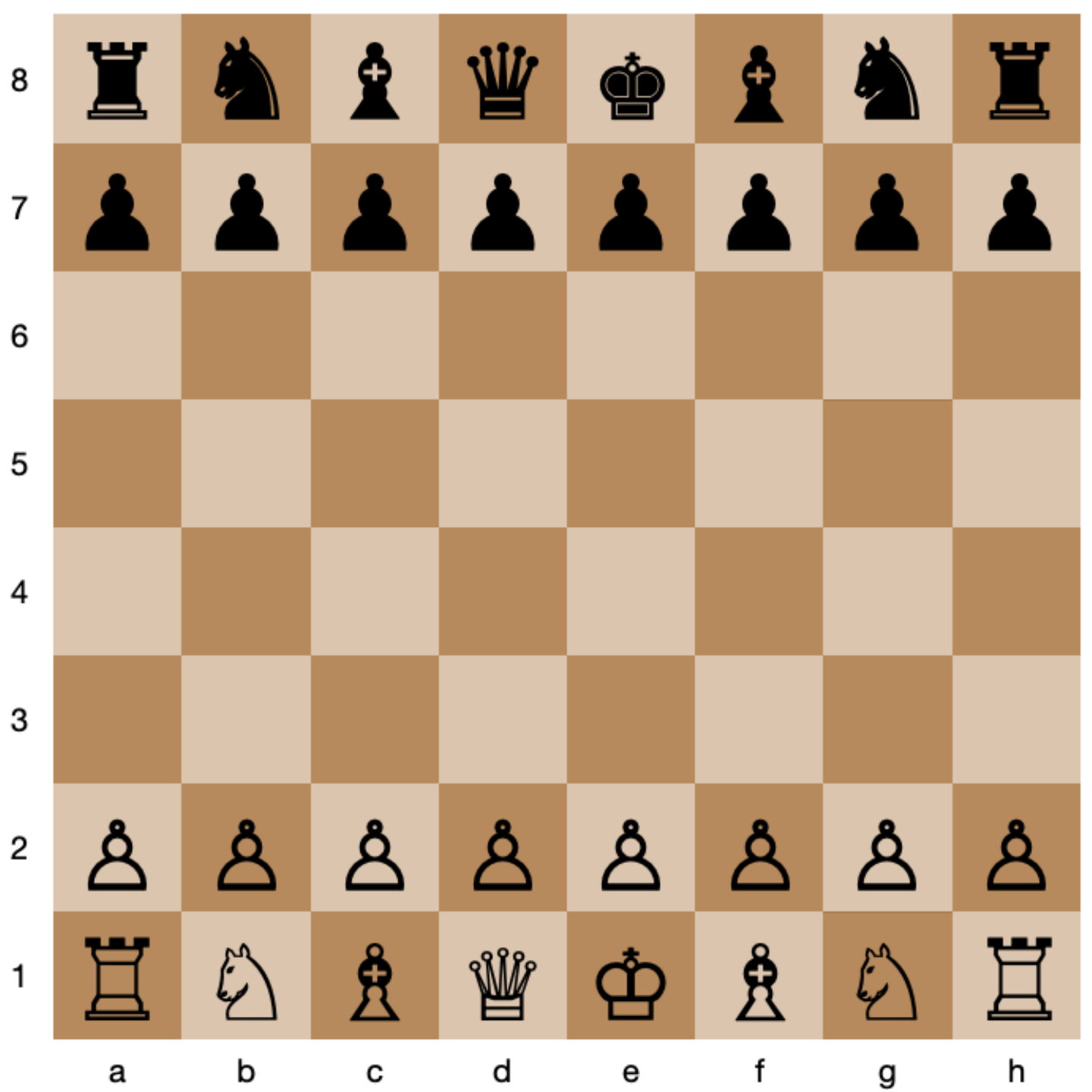}
\caption{Left: conventional chess pieces and their corresponding indefinite pairs with red and blue marks. Right: the initial position of the game is identical to that of conventional chess.}\label{figindef}
\end{figure}

For every conventional chess piece, there is a corresponding pair of 'indefinite pieces', including one piece with a red mark on it and another with a blue one, as shown in Figure~\ref{figindef} (left). The purpose of indefinite pair instances on the chessboard is to represent spatial superposition states of corresponding conventional pieces, as described in the following subsections.\footnote{To differentiate between indefinite pairs of the same type and color, marks with different fill patterns can be used.}

In general, an indefinite piece may move in accordance with the rules for conventional pieces stipulated in~\cite{fide22}. The initial position of the game is identical to that of conventional chess, which means that initially there are no indefinite pieces on the chessboard.

\subsection{Superposition Move}\label{subsec33}

In Niel's Chess, a conventional piece may move to two unoccupied squares simultaneously, or move to one unoccupied square and stay where it is simultaneously.

To execute such a 'superposition move', the conventional piece is replaced by the two pieces of a corresponding indefinite pair instance, placed on the two squares of arrival, one each. The two indefinite pieces are said to be 'paired', as they represent the conventional piece in a superposition state of being (or "spread out") on two squares at once.

\begin{figure}[t]
\centering
\includegraphics[width=0.48\columnwidth]{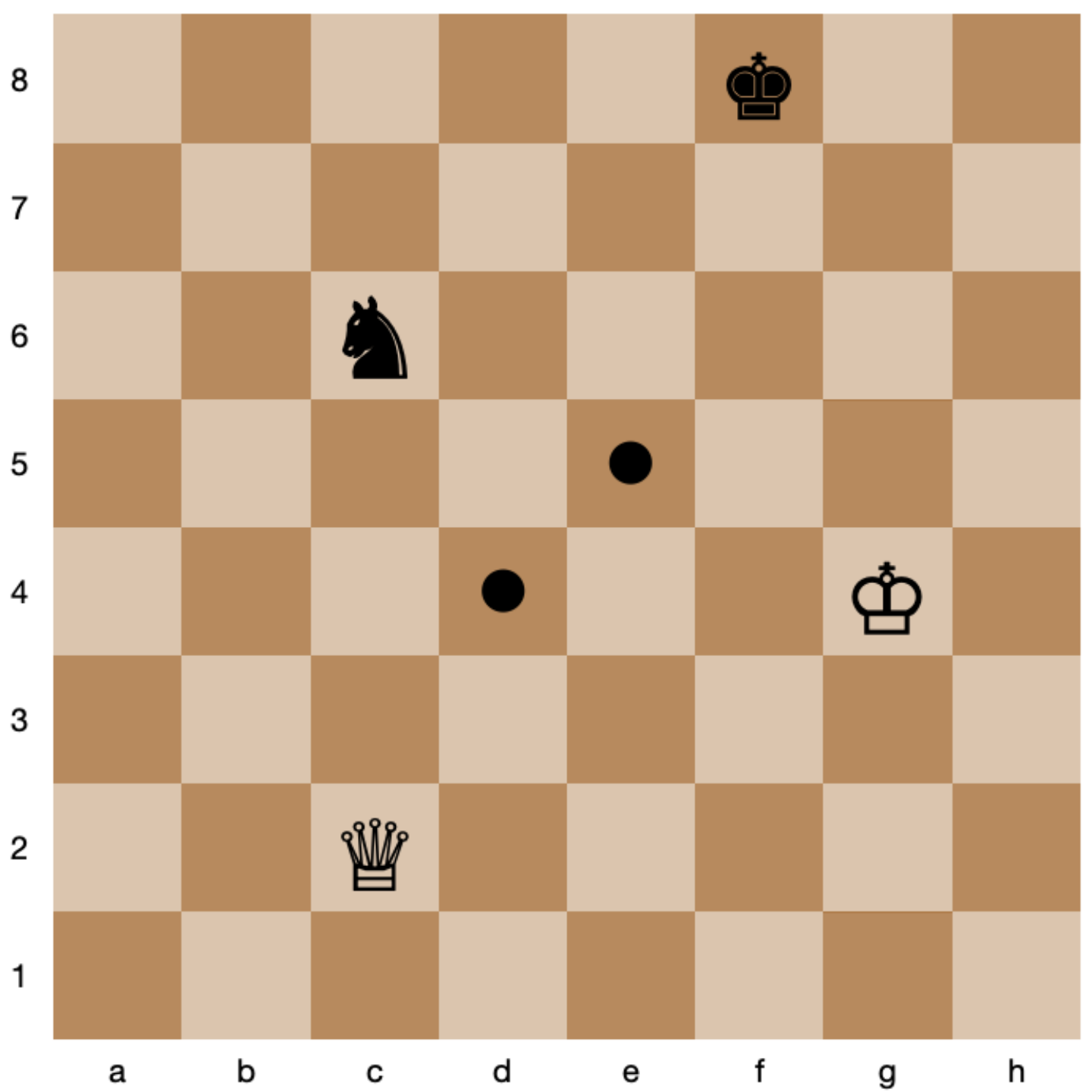}
\caption{The knight on c6 may move to the two unoccupied squares d4 and e5 simultaneously.}\label{figsp}
\end{figure}

\begin{figure}[t]
\centering
\includegraphics[width=0.48\columnwidth]{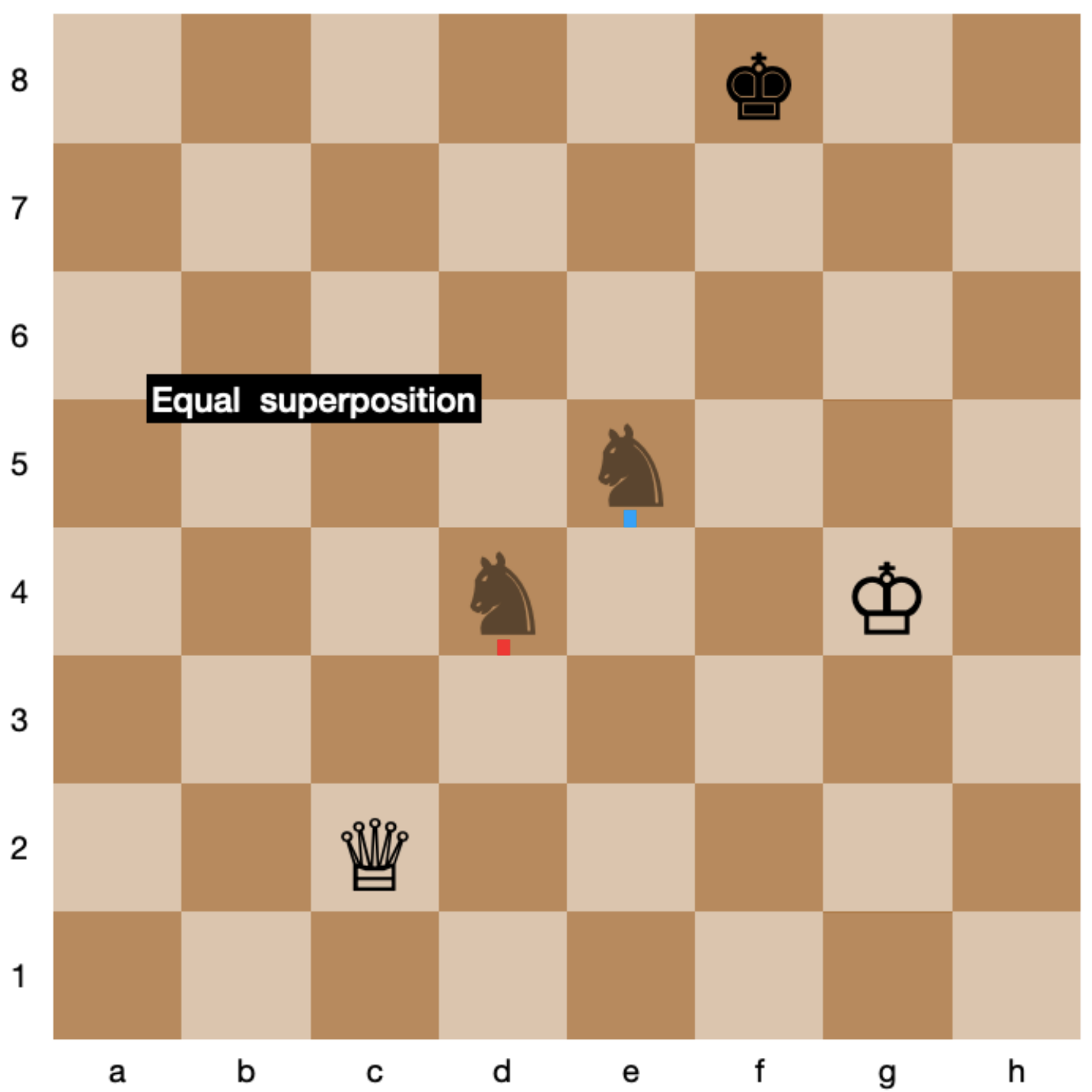}\quad\includegraphics[width=0.48\columnwidth]{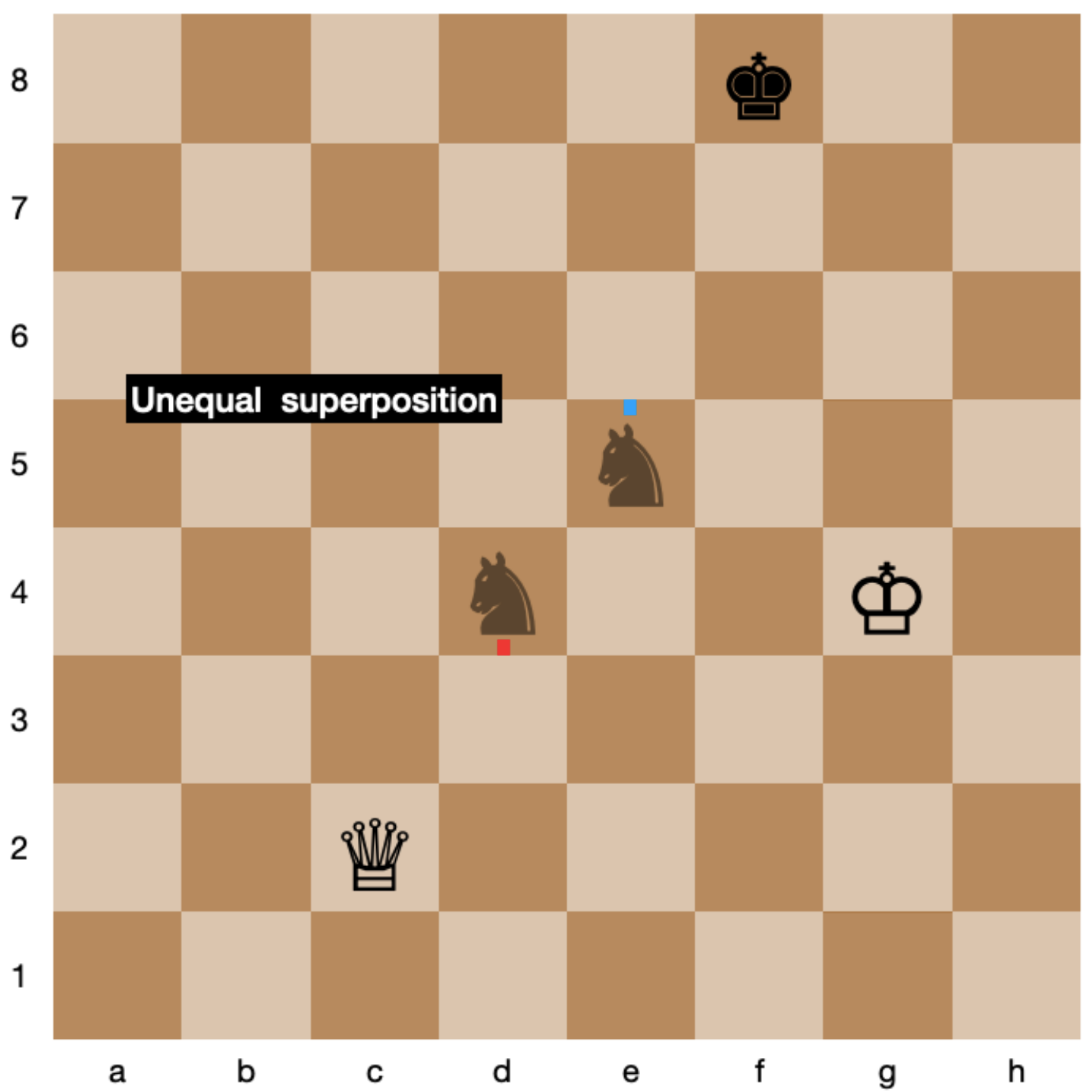}
\caption{Left: equal superposition, the red and blue marks face the same direction. Right: unequal superposition, the marks face opposite directions.}\label{figsm}
\end{figure}

Figure~\ref{figsp} depicts a situation where the knight may simultaneously move to two unoccupied squares, while Figure~\ref{figsm} (left and right) shows two ways of making that superposition move, according to the directions the marks face.\footnote{For better visibility of the directions in 2D, the red and blue marks are displayed separated from their respective indefinite pieces.} If the marks on the paired indefinite pieces both face the opponent, or both face the player who owns the pair, it is called an 'equal superposition'. If one mark faces the opponent but the other the player who owns the pair, it is called an 'unequal superposition'. The player making the superposition move can freely choose either option. No other possibilities are allowed, that is, only facing the upper side or the lower side of the square is permitted.

In order to convey the ideas of the game as clearly as possible, in the following it is assumed that the red mark on an indefinite piece must always face the opponent. This corresponds to the 'Simplified Game' in~\cite{varga23intl}.

\subsection{Attempted Capture}\label{subsec34}

\begin{figure}[t]
\centering
\includegraphics[width=0.48\columnwidth]{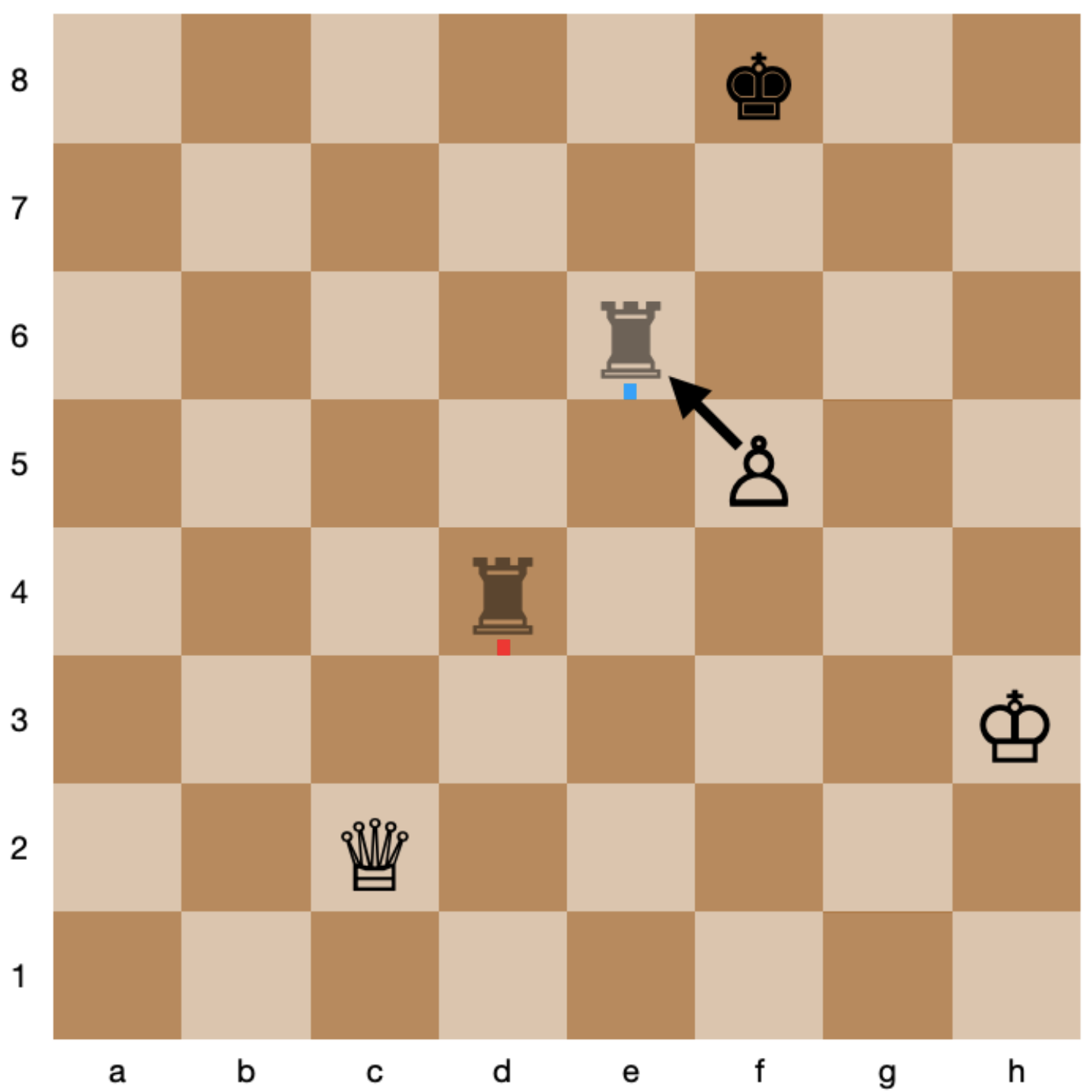}\quad\includegraphics[width=0.48\columnwidth]{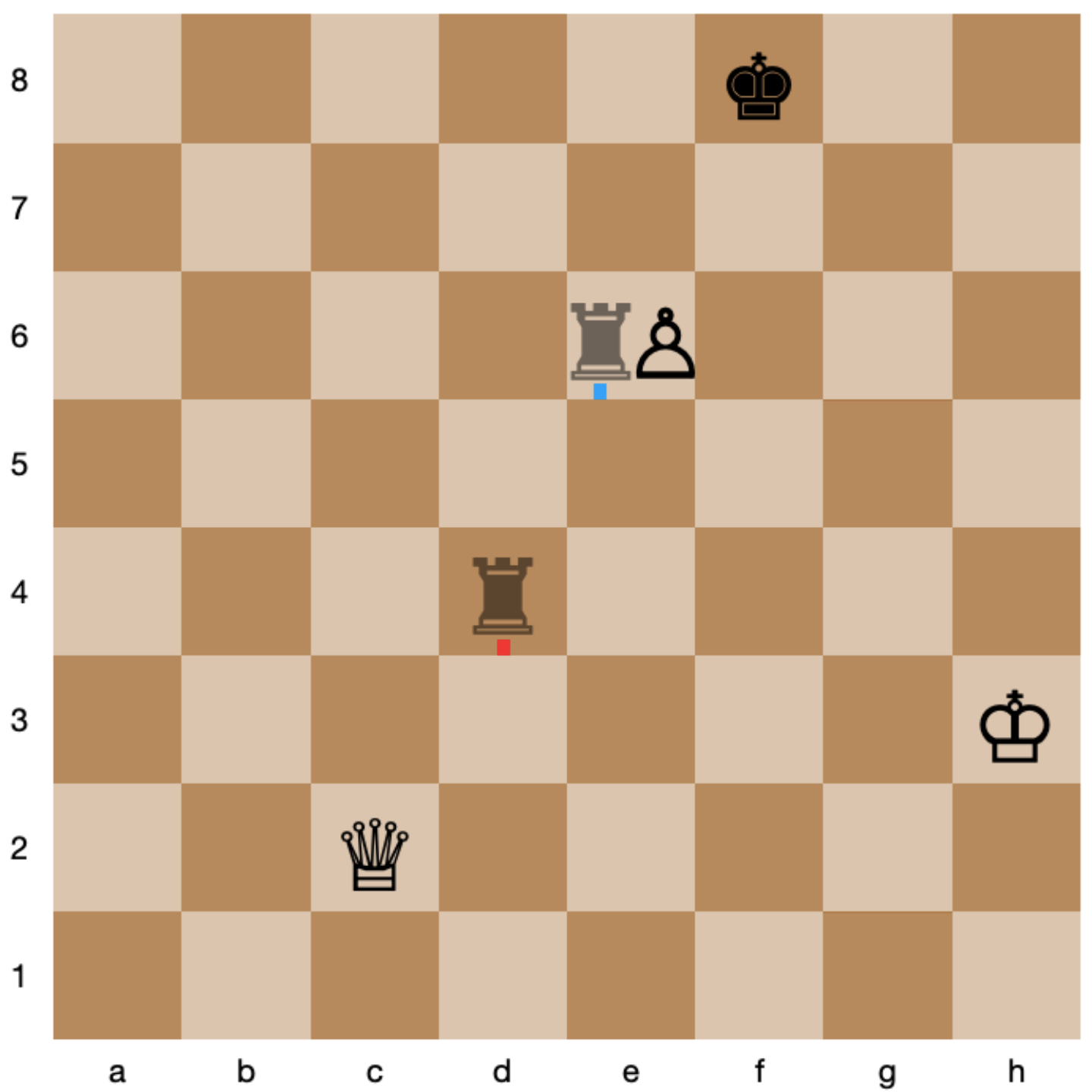}
\caption{Left: the pawn attacks the rook which is on e6 (and d4). Right: the pawn moves to e6 to attempt to capture the rook.}\label{figacp}
\end{figure}

A capture move in which at least one indefinite pair instance is involved is called an 'attempted capture'.\footnote{In other words, either the attacking piece or the targeted piece must be indefinite, or both.} An example is shown in Figure~\ref{figacp}, where the attacking pawn on f5 moves to e6 to capture the opponent's rook. Thus, the first step to execute an attempted capture is to place the attacking piece on the targeted square where the opponent's piece stands. Then, each player who has an indefinite piece on the targeted square randomly collapses their corresponding indefinite pair involved, by rolling a six-sided dice.\footnote{The rationale is that each indefinite piece is thought of as being surrounded by a protection layer (sealed box). However, when it hits, or is hit by, another piece, indefinite or not, the isolation gets broken and the superposition state collapses immediately. A quantum random number generator (QRNG)~\cite{quantis} may be used instead of a dice, to emphasize that the collapse is \textit{truly} random.}

The procedure for randomly collapsing an indefinite pair is as follows. If the pair represents an equal superposition, rolling an even number 2, 4 or 6 (an odd number 1, 3 or 5) 'collapses' the pair to the square where the indefinite piece with the red mark (blue mark) stands. If the pair represents an unequal superposition, rolling a 1, 2, 3 or 4 (a 5 or 6) 'collapses' the pair to the square where the indefinite piece with the red mark (blue mark) stands. As a result of the collapse, the indefinite pair is replaced by the corresponding conventional piece, placed on the square to which the pair has collapsed.

\begin{figure}[b]
\centering
\includegraphics[width=0.48\columnwidth]{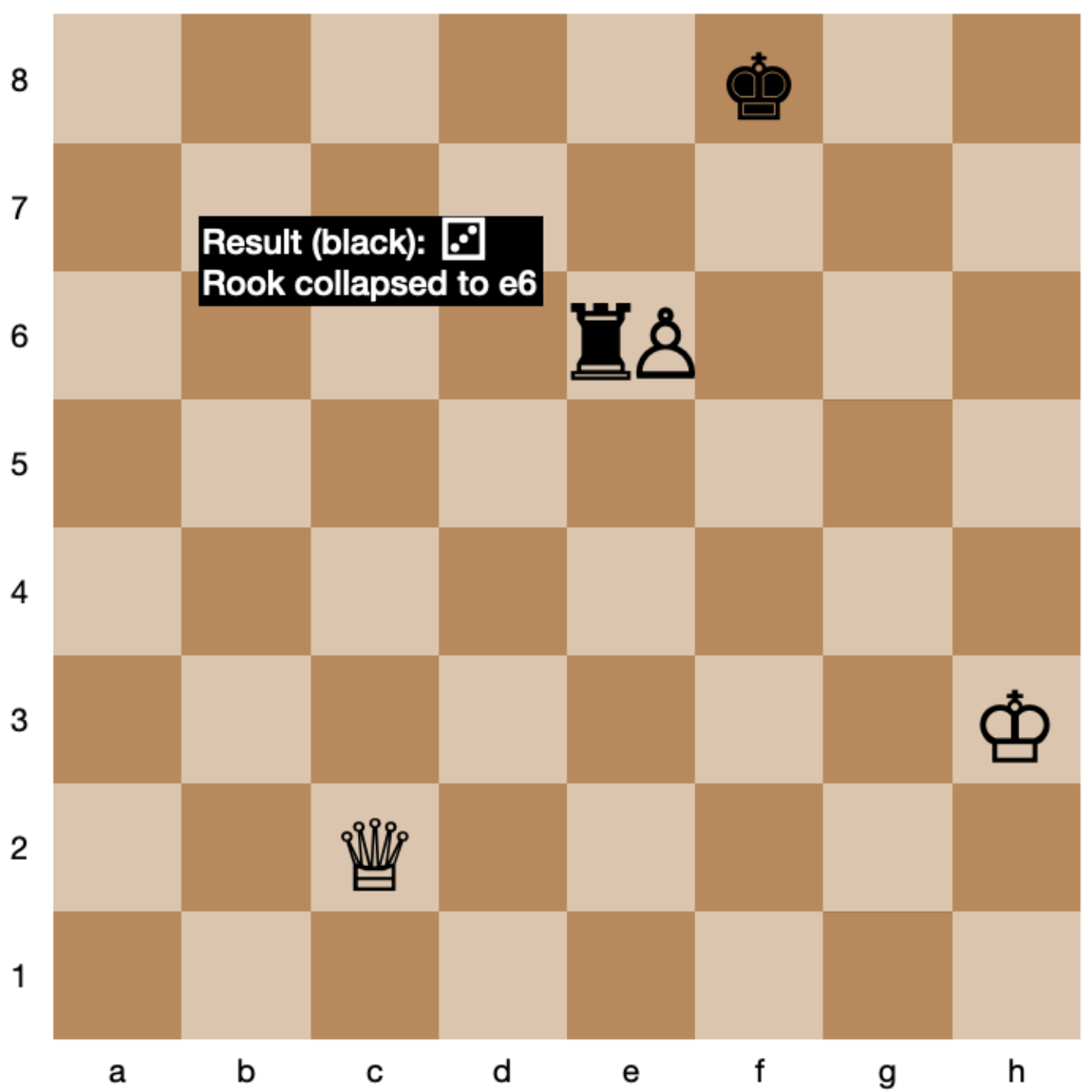}\quad\includegraphics[width=0.48\columnwidth]{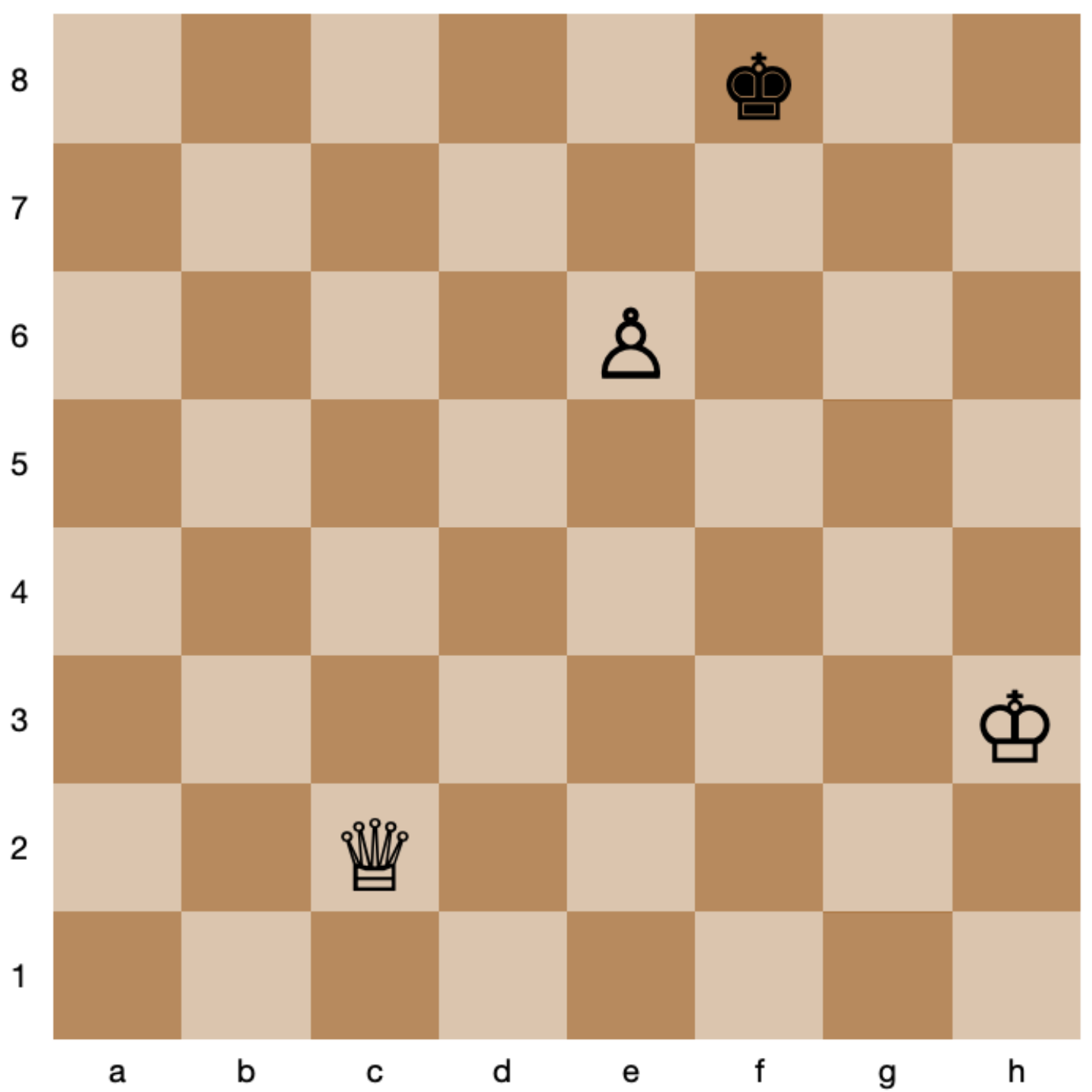}
\caption{Left: the rook collapsed to e6 as Black rolled a 3. Right: the pawn has captured the rook on e6 successfully.}\label{figacs}
\end{figure}

In Figure~\ref{figacp}, the rook is in equal superposition, and then, in Figure~\ref{figacs} (left), Black rolls a 3 which collapses the indefinite pair to e6, that is, where the piece with the blue mark stands. If, after the collapse(s), two conventional pieces end up on the target square, the capture is said to be 'successful': the piece owned by the player attempting the capture stays, while the other is captured and removed from the chessboard, as it happens in Figure~\ref{figacs} (right).

\subsection{Indefinite Check}\label{subsec35}

The king is 'in check' if, in a hypothetical next move, it can either be captured or has a non-zero chance of being captured. For example, the king on g4 in Figure~\ref{figsm} is in check, attacked by the opponent's indefinite knight on e5.

\begin{figure}[t]
\centering
\includegraphics[width=0.48\columnwidth]{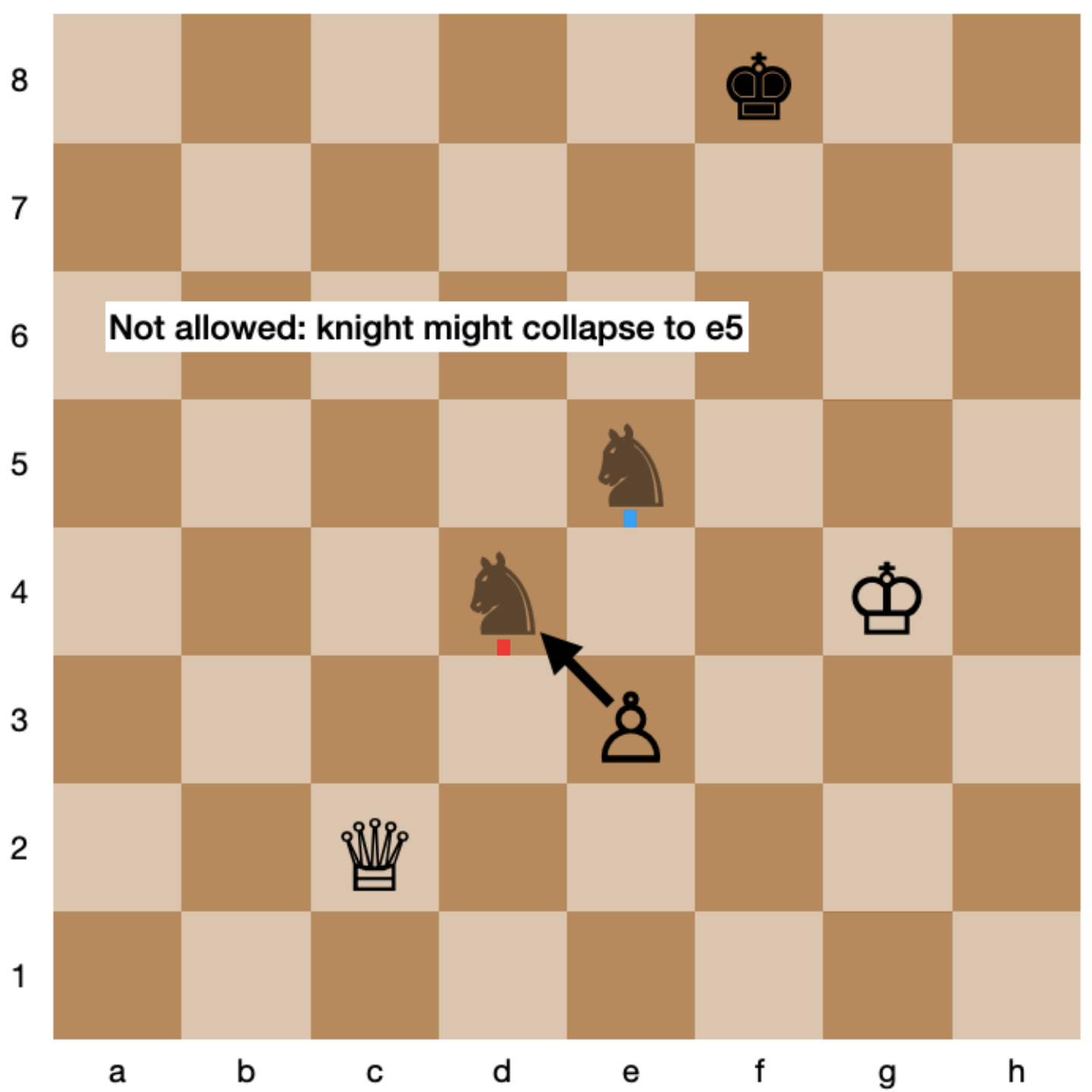}
\caption{The pawn is not allowed to attempt to capture the knight on d4, as the latter might collapse to e5, leaving White's king in check.}\label{figplc}
\end{figure}

Similarly to conventional chess, it is not allowed to make a move that could \textit{potentially} expose the king, indefinite or not, of the same color to check or leave that king in check, by one or more of the opponent's conventional or indefinite pieces. That is, in Figure~\ref{figplc} it is not allowed for White to attempt to capture the knight on d4, as the outcome might be that the knight collapses to e5, potentially leaving the king on g4 in check.

\begin{figure}[t]
\centering
\includegraphics[width=0.48\columnwidth]{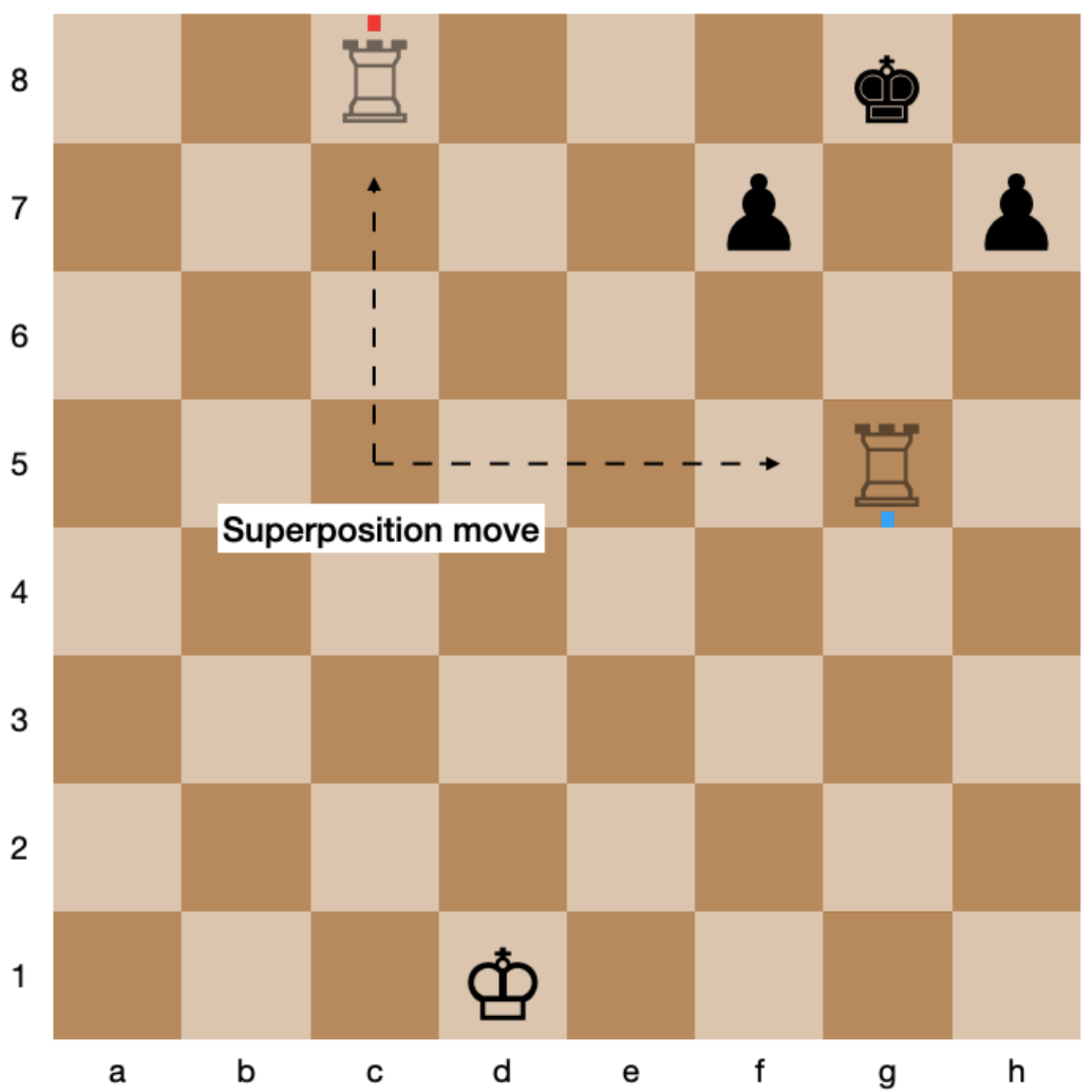}\quad\includegraphics[width=0.48\columnwidth]{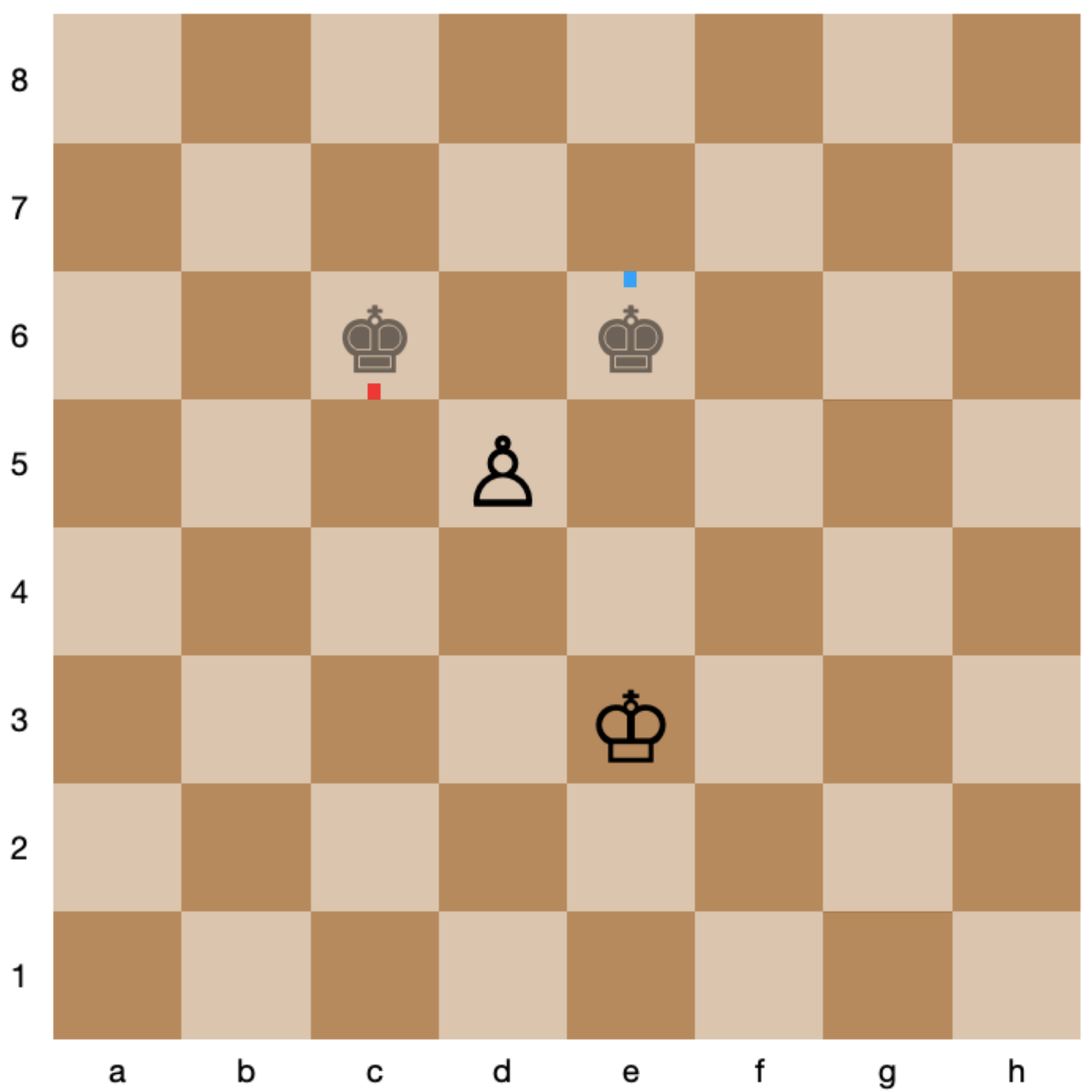}
\caption{Indefinite checkmate examples. The king is in check, and no move can guarantee to get out of it with certainty.}\label{figicm}
\end{figure}

Thus, it is possible to checkmate the opponent's king by placing it under "indefinite attack". Two such examples are shown in Figure~\ref{figicm}. To paraphrase, in Niel's Chess the goal is to place the opponent’s king under risk in a way that no move can guarantee to fully eliminate it.\footnote{One may argue that in the examples of Figure~\ref{figicm}, the king would not be captured with certainty in a hypothetical next move, only with some non-zero probability, and therefore the king should be given a chance to escape, instead of ending the game. Calling such positions a checkmate is inspired by quantum computing. Due to the probabilistic nature of quantum physics (plus the noise in contemporary quantum hardware), it is typical that a quantum program outputs the correct answer to a problem not with certainty but only with some non-negligible probability. And that, provided that the solution can be verified efficiently, already suffices to call the problem "solved".}

\subsection{Entanglement Move}\label{subsec36}

\begin{figure}[t]
\centering
\includegraphics[width=0.48\columnwidth]{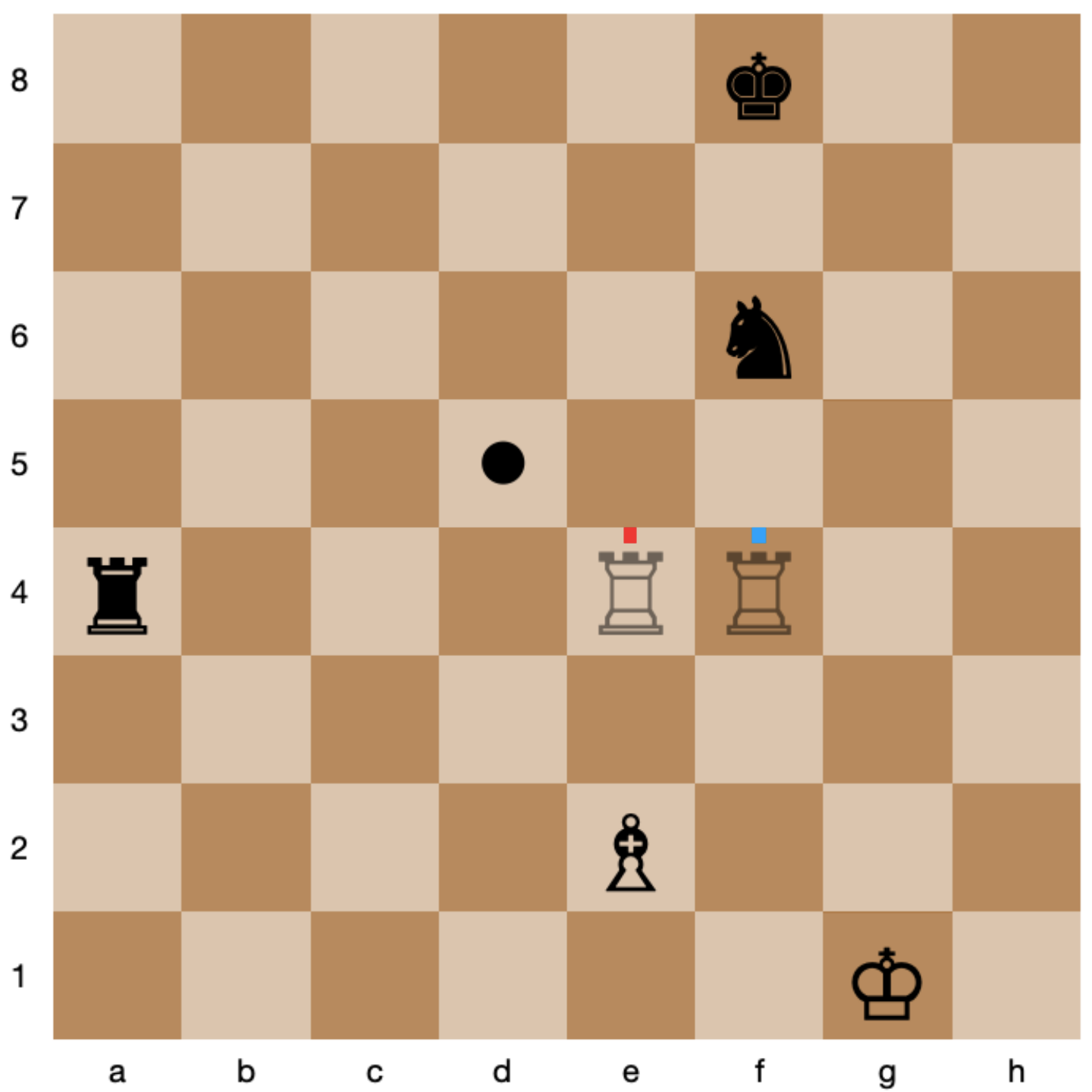}\quad\includegraphics[width=0.48\columnwidth]{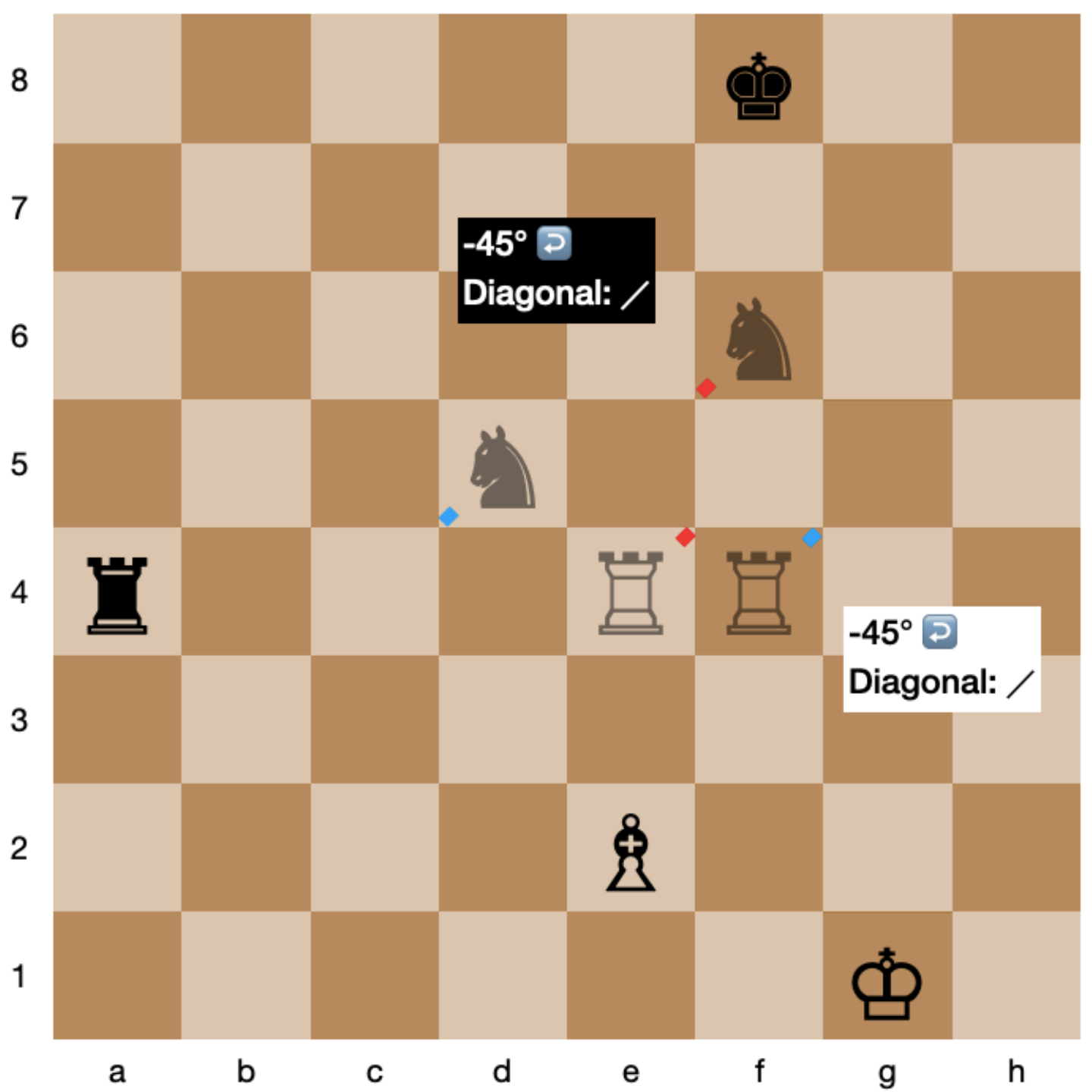}
\caption{Left: knight on f6 may join the (equal) superposition in which White's indefinite rook on e4 participates. Right: after joining, the marks lie on the $-45^\circ$ diagonal, and the knight on f6 attacks the rook on e4.}\label{figentmove}
\end{figure}

\begin{figure}[b]
\centering
\includegraphics[width=0.48\columnwidth]{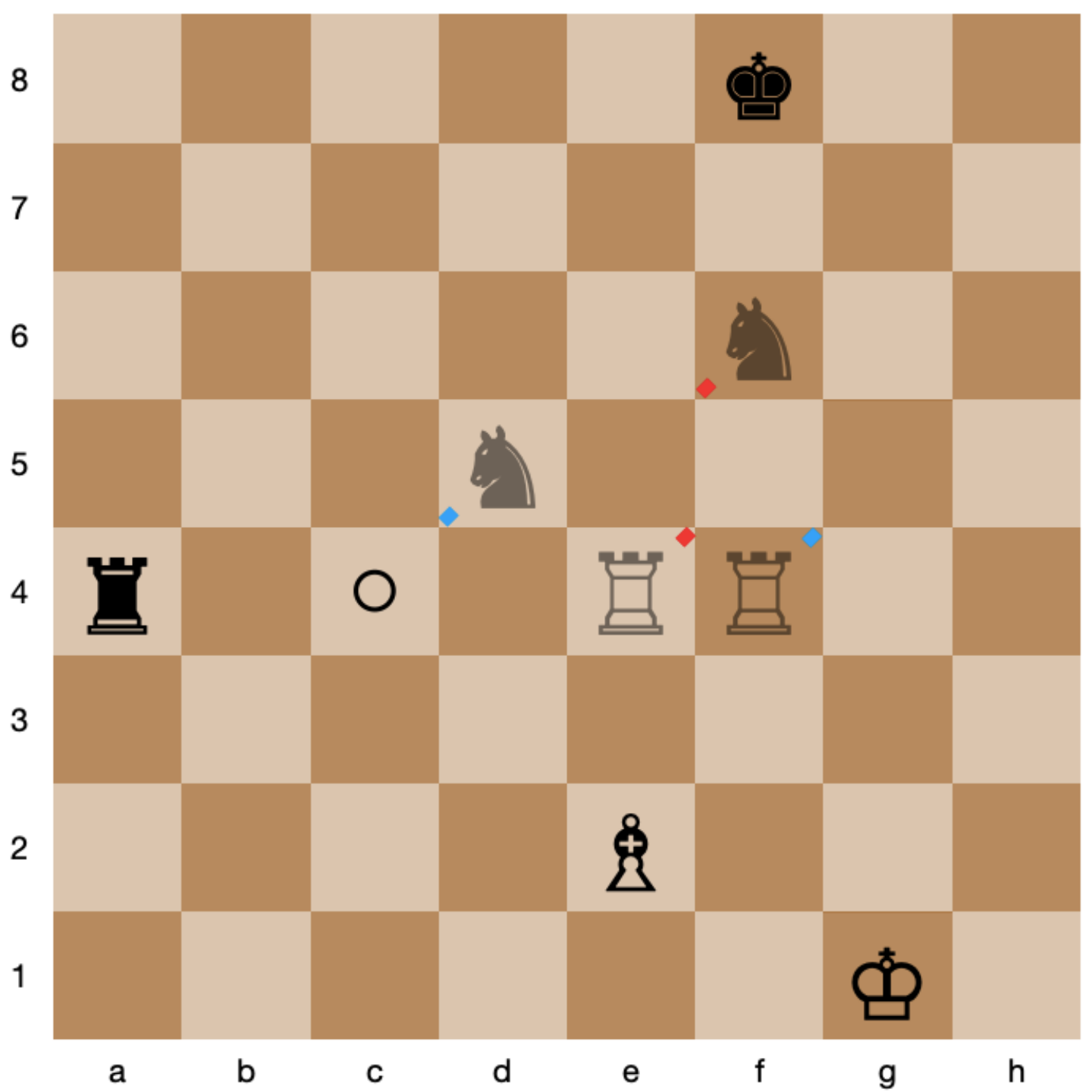}\quad\includegraphics[width=0.48\columnwidth]{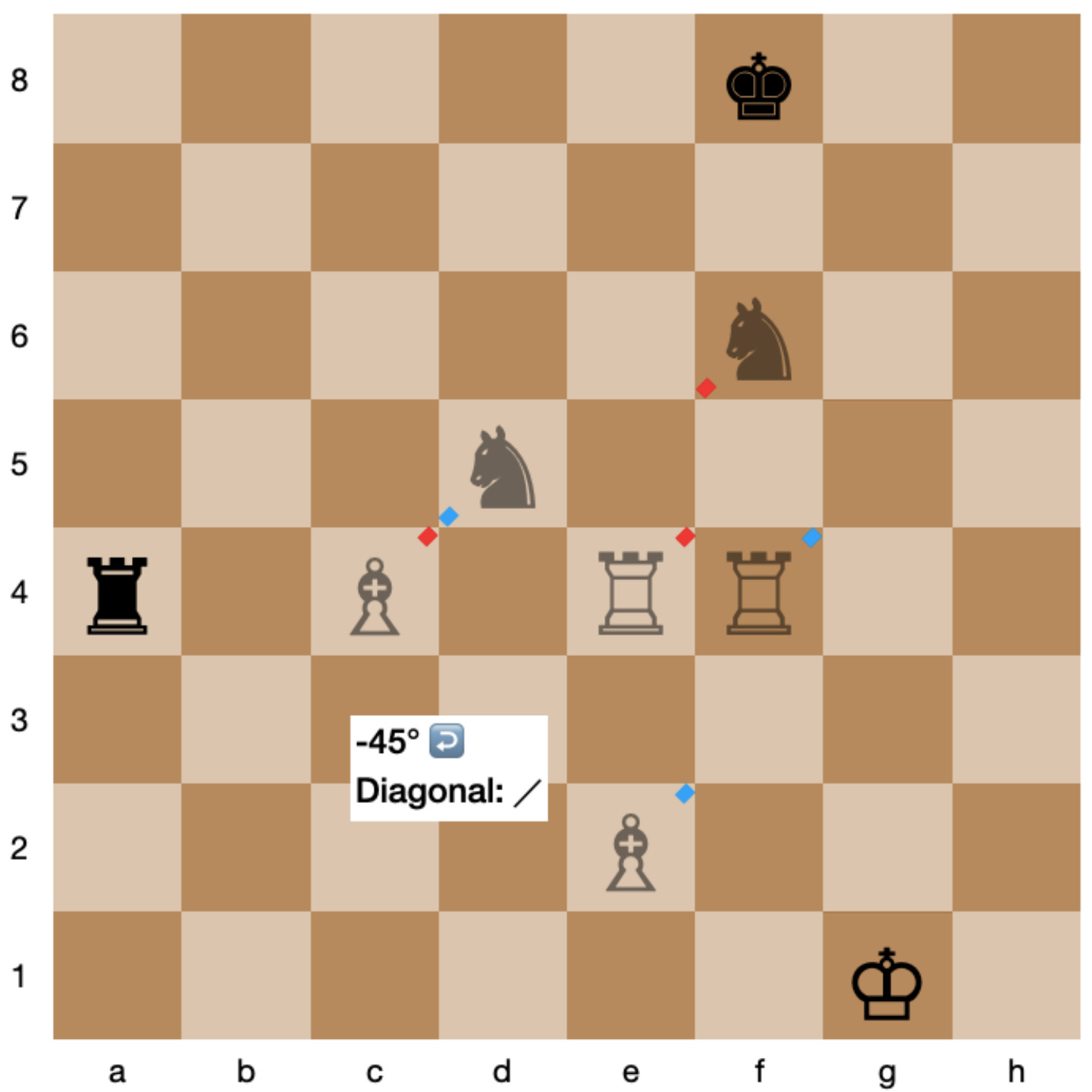}
\caption{Left: the bishop may join as well, by attacking Black's knight from c4. Right: after joining, the marks are aligned.}\label{figentexpansion}
\end{figure}

A conventional piece may join the superposition in which an opponent's indefinite piece participates, by simultaneously moving to an unoccupied square and staying where it is, provided that at least one of the two indefinite pieces which replace the conventional piece attacks the opponent's indefinite piece in question.\footnote{One piece attacking another symbolizes that the latter is "within reach"; direct or indirect proximity of two quantum systems is a prerequisite for creating entanglement between them through operations performed in a lab.} This is called an 'entanglement move'. Figure~\ref{figentmove} (left and right) shows an example where Black's knight joins the superposition in which White's indefinite rook on e4 participates. All the participating indefinite pieces must be oriented such that their marks lie on the same diagonal,\footnote{It might feel uncomfortable that indefinite pieces must be precisely oriented, but this can be taken as a metaphor for quantum states requiring extreme care and precision, compared to dealing with classical states.} and each pair must indicate the same type of superposition (equal or unequal) as the targeted opponent's pair originally indicated.\footnote{Rotating the pieces of an indefinite pair by $45$ degrees does not change the type of superposition, as each mark will face the same upper or lower side of the square as before.} Additional conventional pieces may join later, see Figure~\ref{figentexpansion}.

Indefinite pair instances whose marks lie on the same diagonal are said to be 'entangled'. They represent a corresponding group of conventional pieces in a quantum state of correlated behavior, the details of which are given in Subsection~\ref{subsec37}. Since a square has only two diagonals ($+45$ and $-45$ degrees), there can be at most two entangled groups on the chessboard.\footnote{To justify this limitation, it is noted that in quantum information, entanglement is a \textit{resource}, which can be scarce. That's why, players are allowed only limited access to it, namely, creating new entanglement is not allowed if both diagonals are already taken.}

\subsection{Collapse in Tandem}\label{subsec37}

\begin{figure}[t]
\centering
\includegraphics[width=0.48\columnwidth]{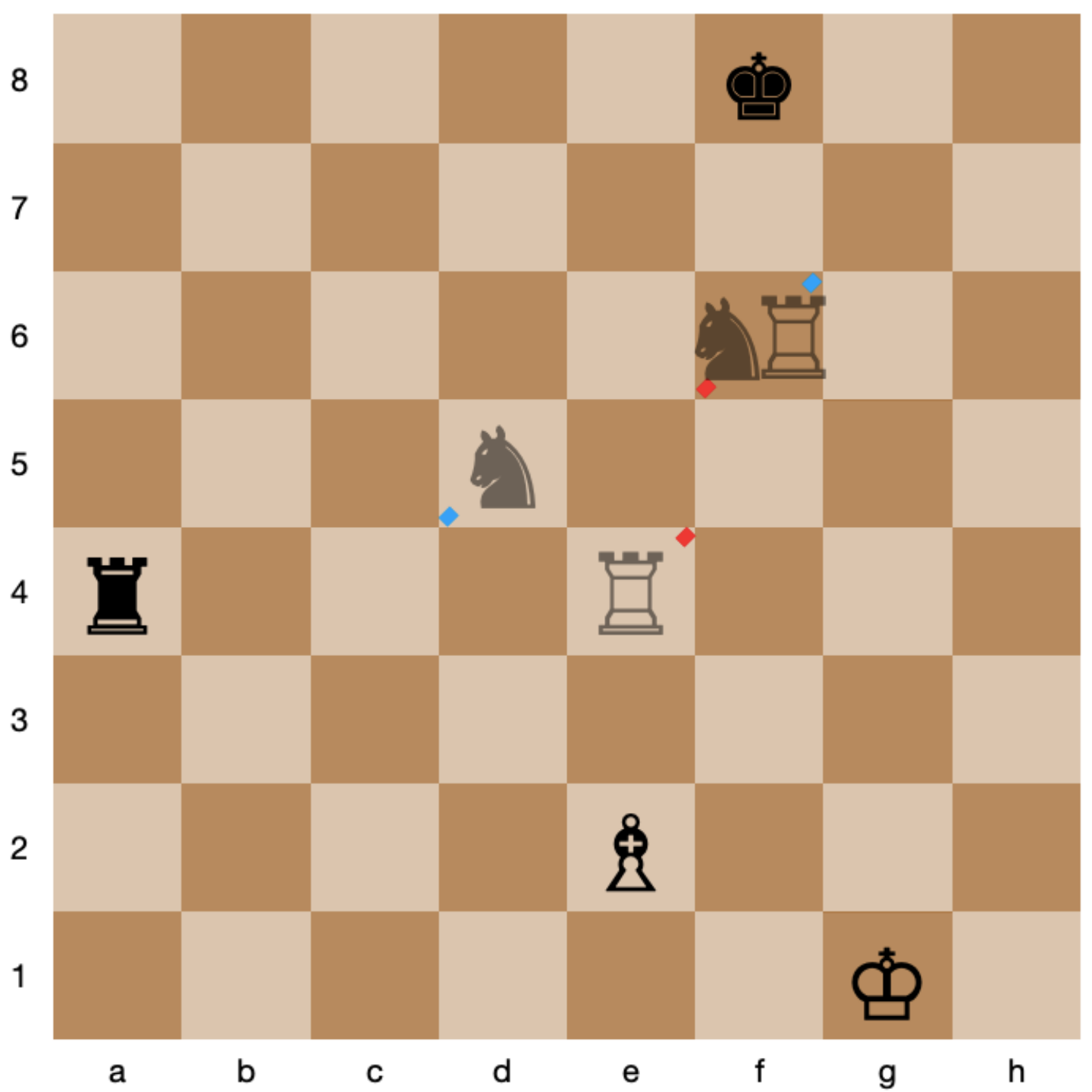}\quad\includegraphics[width=0.48\columnwidth]{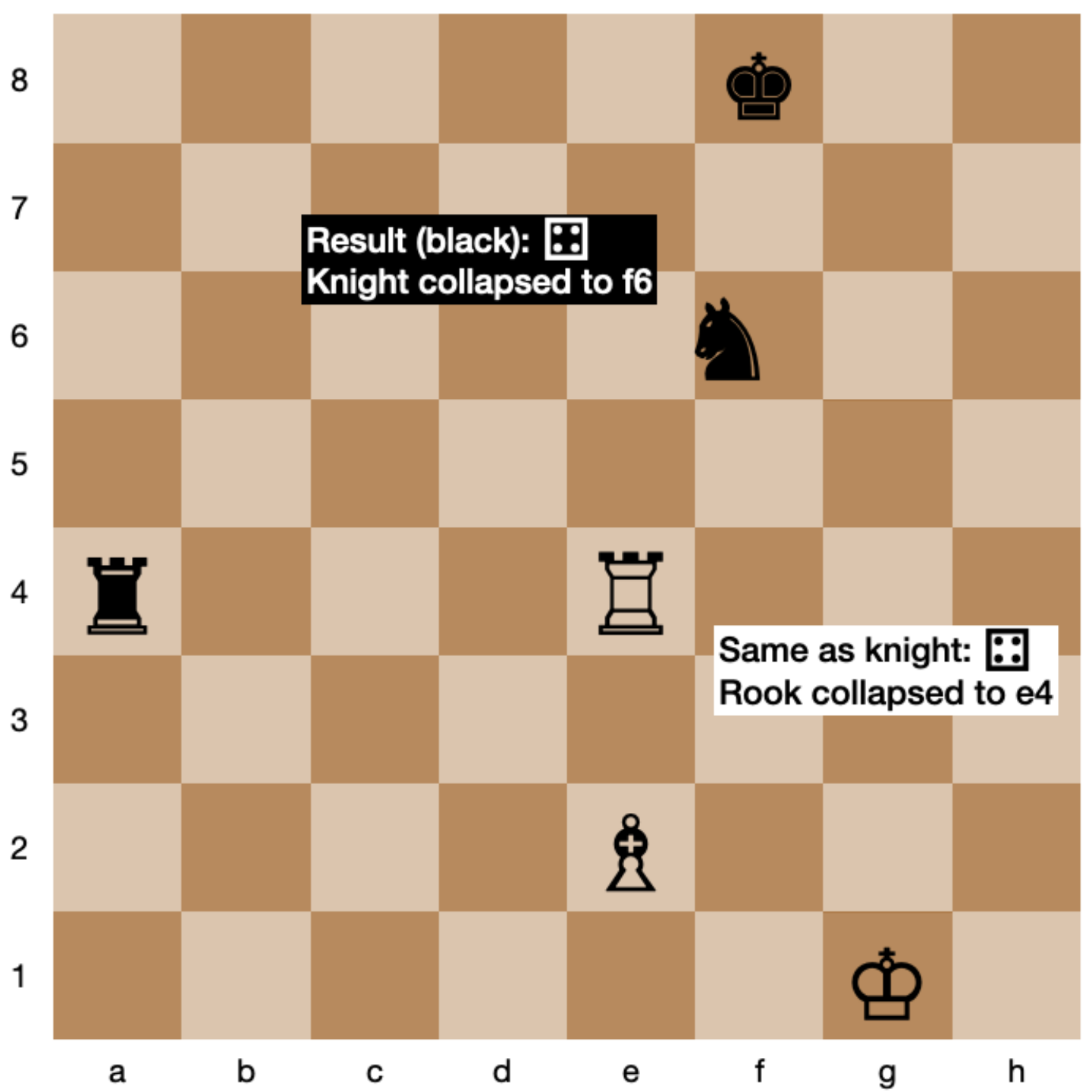}
\caption{Left: rook attempts to capture the knight on f6. Right: knight collapsed to f6 and rook to e4, as Black rolled a 4 (which means collapsing to "red").}\label{figentcollapse}
\end{figure}

Entangled pairs are "connected" in the sense that they collapse in tandem, as illustrated by Figure~\ref{figentcollapse}. That is, collapsing an indefinite pair simultaneously collapses \textit{all} indefinite pairs entangled with it, via a single dice roll. There are two possible outcomes: either every pair in the entangled group collapses to the square with red mark on it, or to the square with blue mark on it.\footnote{Physically, entanglement is a superposition of those two holistic outcomes, "red" and "blue".}

To give another example, let's consider the position on the right side of Figure~\ref{figentexpansion}. If Black's rook on a4 attempts to capture the bishop on c4 (and e2), then all three entangled pairs would collapse, based on the result of White's dice roll to collapse its bishop. However, as the collapse might happen to the "blue" squares, exposing Black's king to check, this move by Black's rook is not allowed.

\begin{figure}[t]
\centering
\includegraphics[width=0.48\columnwidth]{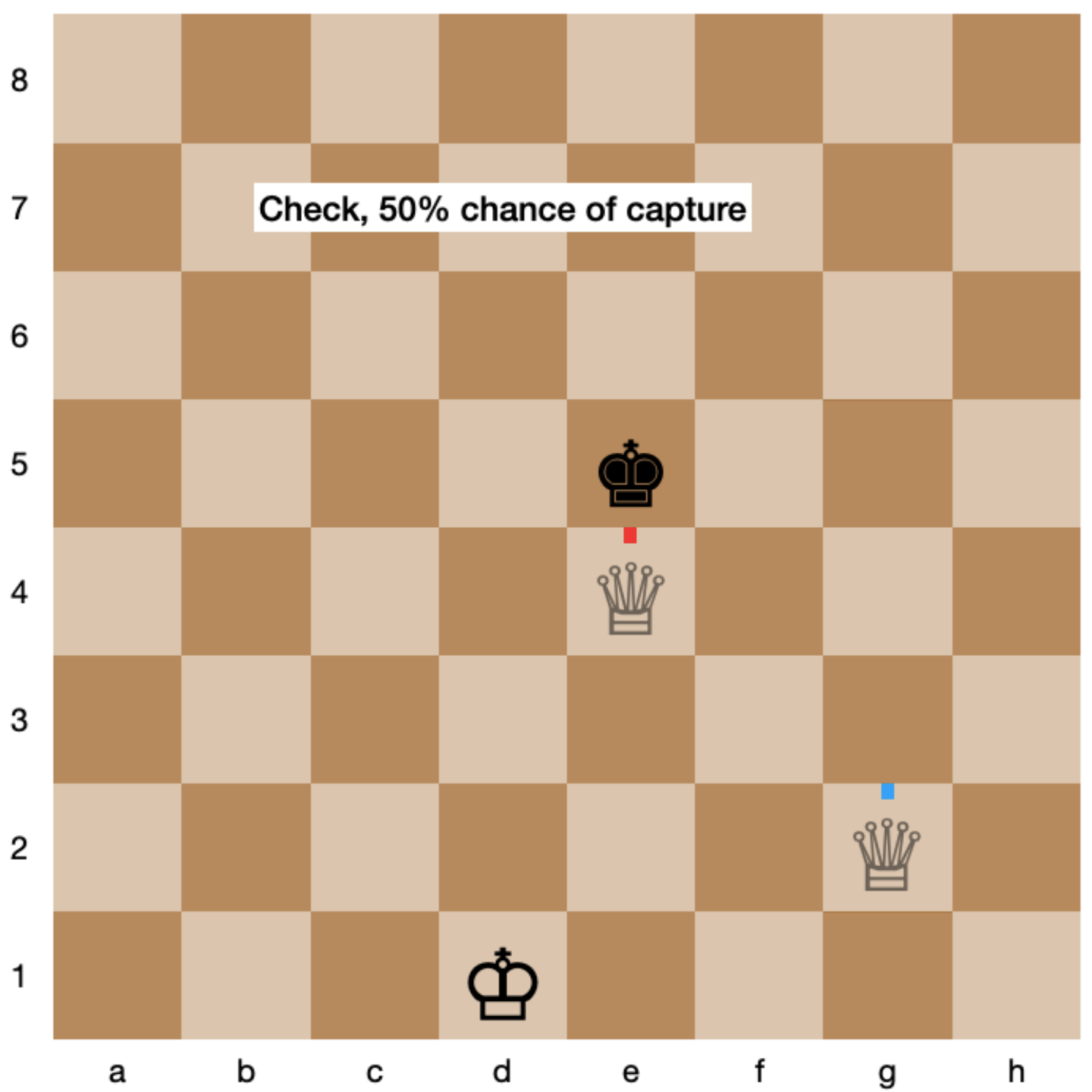}\quad\includegraphics[width=0.48\columnwidth]{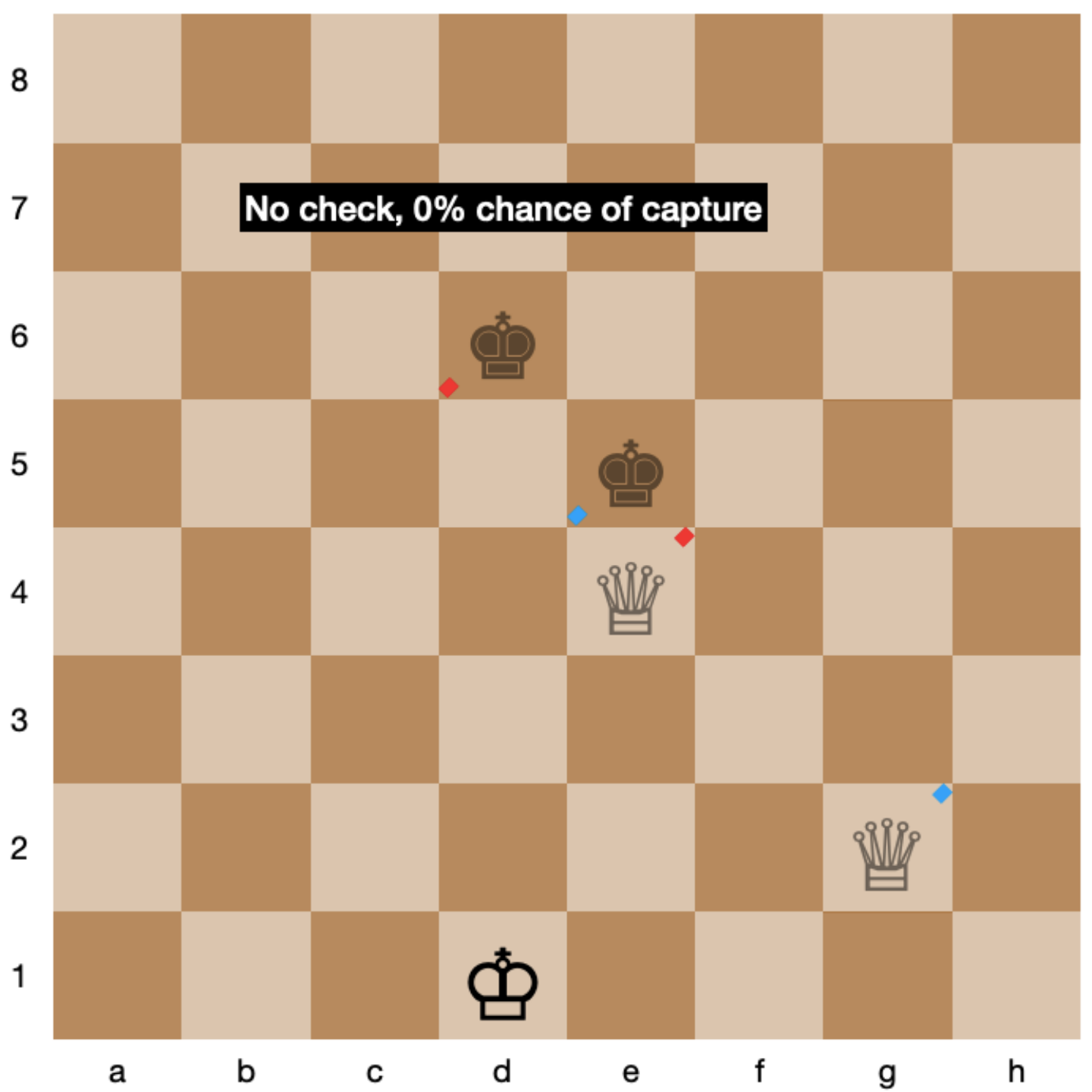}
\caption{The king on e5 is in check (left), but Black is allowed to entangle its king with White's queen to get out of check (right).}\label{figzerorisk}
\end{figure}

Finally, entanglement gives rise to a strange situation: if the king, indefinite or not, is attacked by an opponent’s piece such that a hypothetical capture of that king would have zero chance of success, the attack does not count as a check. See an example in Figure~\ref{figzerorisk}.

\subsection{Phase and Interference}\label{subsec38}

As an alternative to unequal superposition, the red and blue marks facing opposite directions could be interpreted as an 'equal superposition with (a $180$-degree) relative phase'~\cite{varga23intl}. One can work out such a variant of Niel's Chess to demonstrate that quantumness is richer than mere statistical uncertainty,\footnote{Although it is already remarkable, even without considering phase, that we have \textit{true} randomness, not just ignorance; and also that entanglement exhibits a special kind of (connected) uncertainty.} introducing moves analogous to the Pauli-$Z$ gate (via a $180$-degree rotation of an indefinite piece with a blue mark on it) and the Hadamard gate. See the 'Hadamard Game' in~\cite{varga23intl}.

\section{Pilot Experiment}\label{sec4}

To investigate whether young children can learn how to play Niel's Chess, the author conducted a pilot education session with four occasional chess players: two 10-year-olds, an 11-year-old, and a 12-year-old. All four children were previously acquainted with the author. In addition to learning how to play the game, the goal was to provide the children with a playful experience about key concepts in quantum physics, to raise their interest and motivation to explore the subject further.

Since the session was planned for only 2 to 2.5 hours, it did not include all the aforementioned ideas and rules of Niel's Chess. The included quantum-physical concepts were equal superposition, random collapse, and quantum tunneling (see Figures~\ref{figtunnel1} and~\ref{figtunnel2} in Appendix~\ref{secC}). The corresponding set of rules already make up an enjoyable and challenging logic game. The left-out topics of unequal superposition and entanglement can be covered in a second session, while those of phase and interference in a third one, by gradually introducing new rules to the game.

In order to maintain the attention of the students, the event was set up as a competition between two teams of two, whereby the teams collected points by answering questions, solving chess puzzles, and eventually playing Niel's Chess against each other. In the end, both teams received prizes. The language of instruction was German, and the pilot was structured as follows:

\textbf{Quantum is coming.} Participants were told that Niel's Chess is a game of the future, due to the expectation that quantum information technologies will become widespread in about a decade, during their time in tertiary education. Thus, quantum thinking will be a crucial skill in creating innovative applications.

\begin{figure}[t]
\centering
\includegraphics[width=0.96\columnwidth]{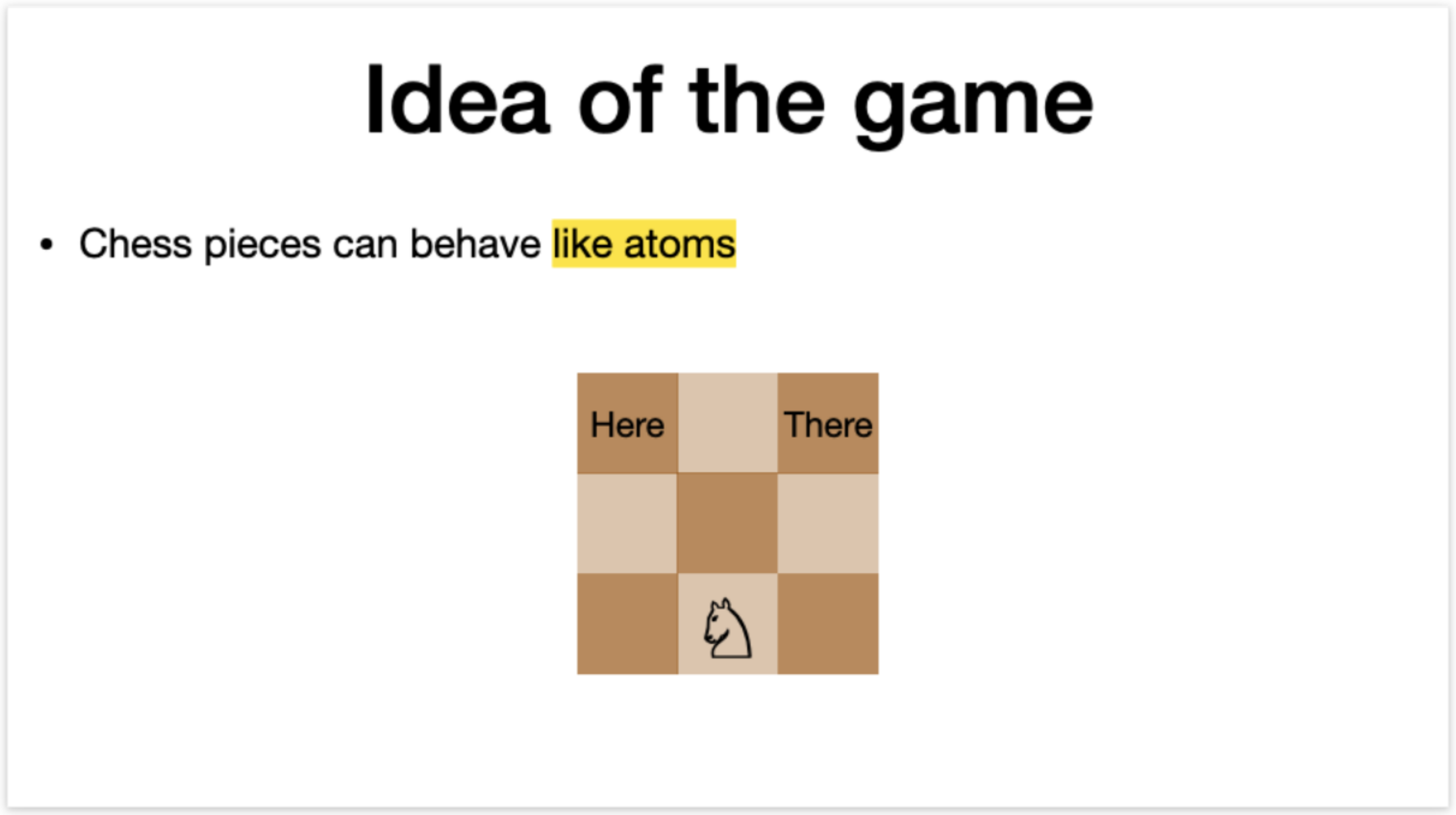}
\caption{The knight may move to both squares "Here" and "There" simultaneously, attaining a state of spatial superposition.}\label{figidea}
\end{figure}

\textbf{The world of atoms.} Quantum physics was introduced as the theory describing the weird laws governing the world of atoms and elementary particles. According to these laws, atoms can be in spatial superposition. Then, Niel's Chess was framed as a game in which chess pieces can behave like atoms, see the slide (translated into English) in Figure~\ref{figidea}.

\begin{figure}[t]
\centering
\includegraphics[width=0.96\columnwidth]{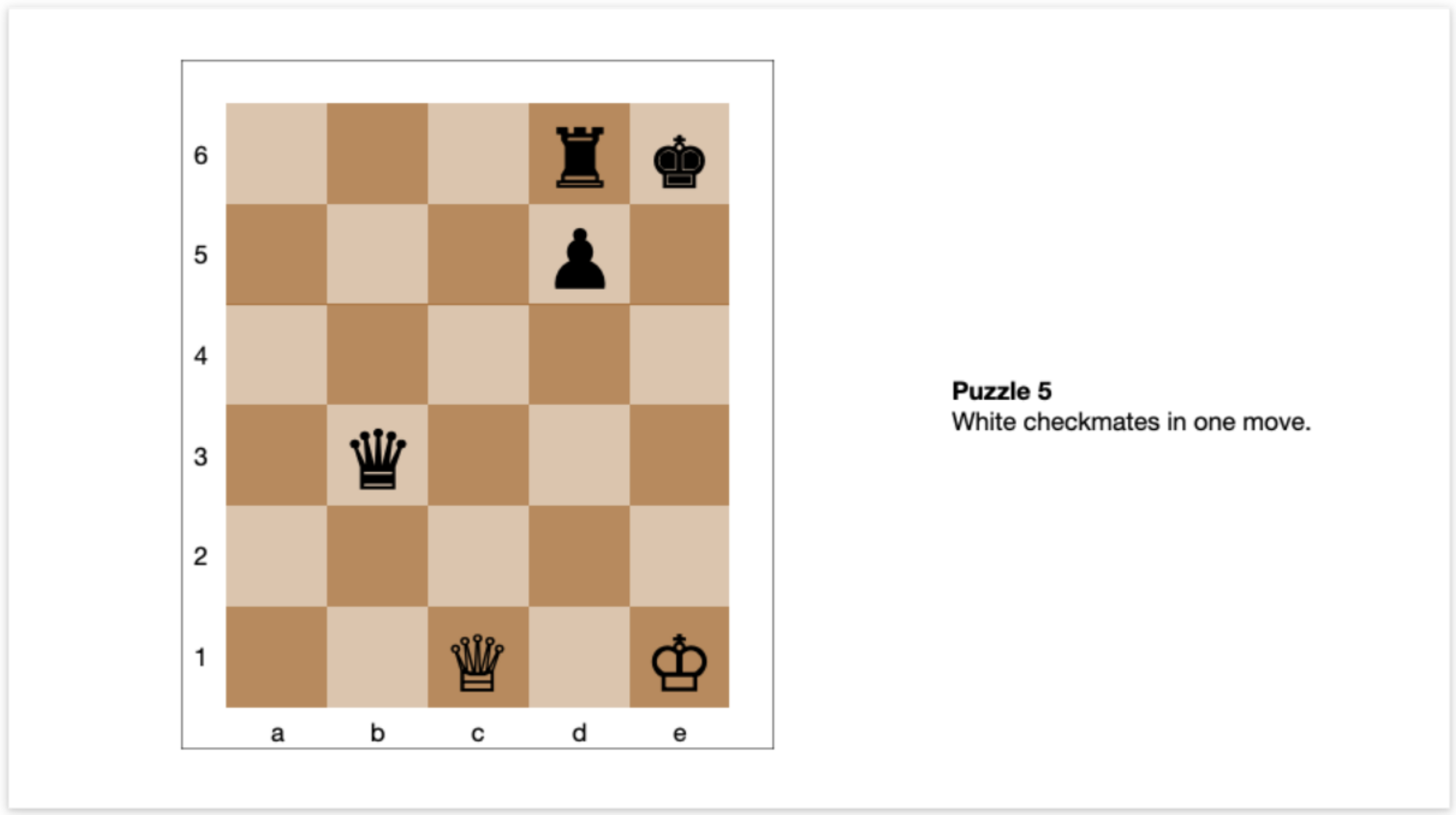}
\caption{White's queen can checkmate Black's king by moving from c1 simultaneously to c3 and e3.}\label{figpuzzle}
\end{figure}

\textbf{Rules of the game.} The equal superposition move, attempted capture, indefinite check, and quantum tunneling were explained and internalized by solving chess puzzles. A $5\times 6$ board was used throughout the session, so as not to overwhelm the children with complexity. The puzzles were intended to be simple, but still instructive and challenging enough. One example is shown in Figure~\ref{figpuzzle}.

\textbf{Schrödinger's Cat.} The possibility of macroscopic superposition was touched upon, starting with the observation that everyday objects consist of atoms, which individually obey the laws of quantum physics. Then, continuing with the Schrödinger's Cat thought experiment~\cite{schroedinger35}, it was pointed out that superposition states must be extremely well isolated from the environment to prevent them from collapsing. Finally, the idea of non-spatial superposition was illustrated by an analogy in which a glass of water is simultaneously hot and cold.

\textbf{Minichess game.} The last 30 minutes of the session were spent playing Niel's Chess on a $5\times 6$ board, two games in parallel by opposing team members, using the rules learned earlier. The initial position of the game can be seen in Figure~\ref{figmini} in Appendix~\ref{secC}.

\section{Results and Discussion}\label{sec5}

The children were active and remained focused throughout the entire event; no break was needed at all. They found the prospects of macroscopic superposition (Schrödinger's Cat) especially intriguing. Moreover, they were highly motivated in the competition, eager to find a solution to every single task. And, most importantly, they had no problem applying the quantum rules when playing Niel's Chess. It was observed though that they tended to make superposition moves just for the sake of trying, rather than having a concrete plan. This is understandable, as it is a common experience that quantum principles take time to assimilate.

Regarding the competition, the teams had 4 questions to answer about quantum physics during the session, complemented by 8 chess puzzles involving superposition and quantum tunneling. Each of these 12 tasks could be solved correctly by at least one team, while 7 of them were solved by both teams. It is important to note that the purpose of these tasks was to gain understanding rather than to test it. Giving hints was part of the puzzles, because none of the four children are frequent chess players, and they had got to know the quantum rules literally minutes before. The hints led them to search for the solution in terms of a corresponding pattern of the game. For example, the hint for the puzzle shown in Figure~\ref{figpuzzle} was the single word "blocking", based on which both teams found the solution quickly.

Altogether, the pilot experiment suggests that children as young as 10 years old can learn how to play Niel's Chess, exposing them to some of the most important principles and thinking patterns of the quantum world at an early age. As for older children, a recent study shows that high school students already possess all the necessary background to learn rigorous, comprehensive quantum theory~\cite{coecke23}. For this latter age group, Niel's Chess can supplement other, curricular materials for quantum technology education.

Besides schools, Niel's Chess can also be an ideal outreach tool, targeting those people in the general public who know the rules of chess but do not wish to delve into rigorous quantum physics. Within the community of players, the game can be used as a starting point to discuss quantum principles and clear up possible misconceptions.

\section{Conclusion}\label{sec6}

In this paper, a quantum variant of chess called Niel's Chess was presented, which can be played on a traditional board, without using computers or other electronic devices. For the younger generations, it aims to help students develop problem solving skills for the uncertain and interconnected world of today, and prepare them for the world of tomorrow shaped by quantum technologies.

A pilot experiment suggests that the game can potentially be used to introduce foundational principles of quantum physics already to children from the age of 10. The same methodology of marks and rotations can be applied to make other board games "quantum" as well, such as Chinese Chess (Xiangqi)~\cite{varga23chn}, allowing educators to engage perhaps even younger children, by selecting games suitable for the given age group, country and culture.

\section*{Acknowledgment}

The author thanks Manuel Oriol and Yaiza Aragon\'es-Soria for their helpful feedback and discussions, and the anonymous reviewers for their valuable comments and observations.

\section*{Declarations}

Niel's Chess is a trademark co-owned by the author, and there is a related patent pending in multiple jurisdictions. The teaching materials (in German) used in the pilot are available from the author on reasonable request.

\begin{appendices}

\section{The Chess Set}\label{secA}

The quantum chess set can be produced easily in a DIY manner, by combining (and marking) the pieces of three conventional chess sets. For each conventional piece, there are two corresponding indefinite pieces of the same size, shape and color, one with a red mark and the other with blue. Ideally, the indefinite pieces are either of a more matte or transparent finish, to be able to better differentiate them. However, they may also be of the same finish as the conventional pieces.

Alternatively, a single conventional chess set would suffice if each conventional piece consists of two easily separable halves, held together, for instance, by magnetic force. The halves would then play the role of the corresponding indefinite pieces, one with a red mark on it, the other blue.

\section{Piece in a Box}\label{secB}

Both Niel's Chess and Cantwell's quantum chess~\cite{cantwell19} are based on the spatial superposition of conventional chess pieces. However, there is an important conceptual difference between the two approaches, which can be illuminated in terms of the Schrödinger's Cat thought experiment~\cite{schroedinger35}.

The underlying idea in Niel's Chess is “piece in a box”, meaning that each indefinite piece is analogous to half of the sealed box in a Schrödinger's Cat-like setup. In this setup, instead of being poisoned or not, the cat truly randomly chooses which half of the box to stay in, and then the box is \textit{very carefully} separated into two sealed half-boxes. Later, whenever such a sealed half-box hits, or is hit by, a conventional piece or another sealed half-box, the isolation gets broken and the superposition state inside the whole box collapses immediately.\footnote{For completeness, recall that in Akl's quantum chess~\cite{akl16} it is the player's hand touching the piece that causes the (non-spatial) superposition to collapse.}

In contrast, the game in~\cite{cantwell19} evolves as a single, overall superposition of purely conventional positions on the chessboard. There are no "indefinite pieces", since no piece is isolated individually. It is rather as if the whole chessboard, including the conventional pieces on it, had been placed into a sealed box at the beginning of the game. This idea might be called “chessboard in a box”. In such a setup, no collapse would be necessary per se, because according to quantum theory, informational isolation does not have to be broken in order to move the pieces inside the box via applying unitary operations~\cite{schumacher10}. Nevertheless, (partial) collapses do happen in the game, targeting one or more squares, triggered by the ”system” due to pragmatic reasons, to keep the complexity of the superposition state inside the box manageable.

The distinct underlying physical viewpoints, that is, "piece in a box" vs. "chessboard in a box", give rise to characteristic differences between Niel's Chess and Cantwell's quantum chess, especially in the way attempted captures and entanglement moves are executed. These are complemented by several additional differences in gameplay, such as the meaning of check and checkmate, unequal superposition and the swapping of collapse probabilities thereof (see in~\cite{varga23intl}), and that Niel's Chess can be played without relying on a computer. This last feature makes the process of experimenting with new rules easier for educators.

\section{Variants of the Game}\label{secC}

\begin{figure}[t]
\centering
\includegraphics[width=0.23\columnwidth]{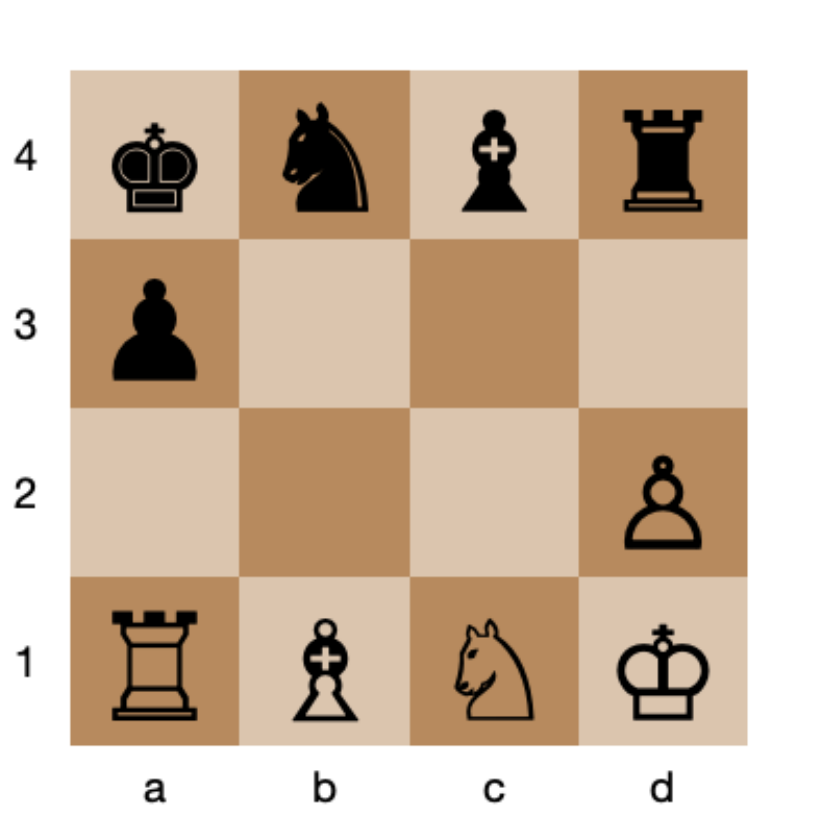}\;\includegraphics[width=0.23\columnwidth]{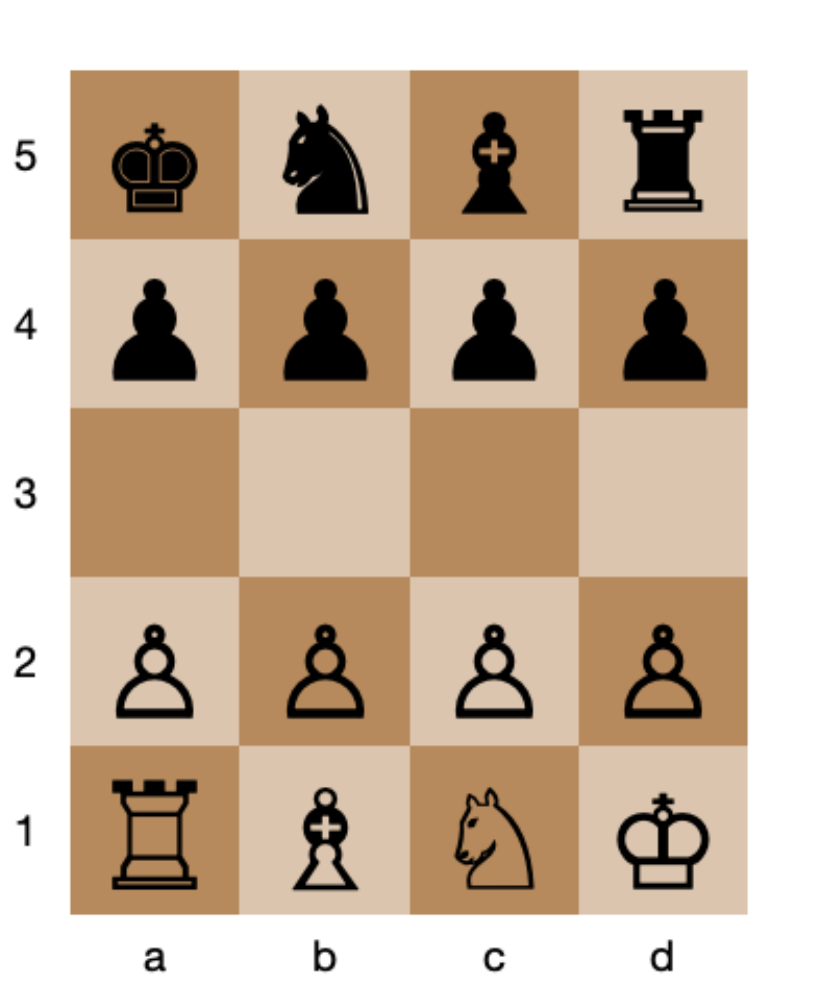}\;\includegraphics[width=0.23\columnwidth]{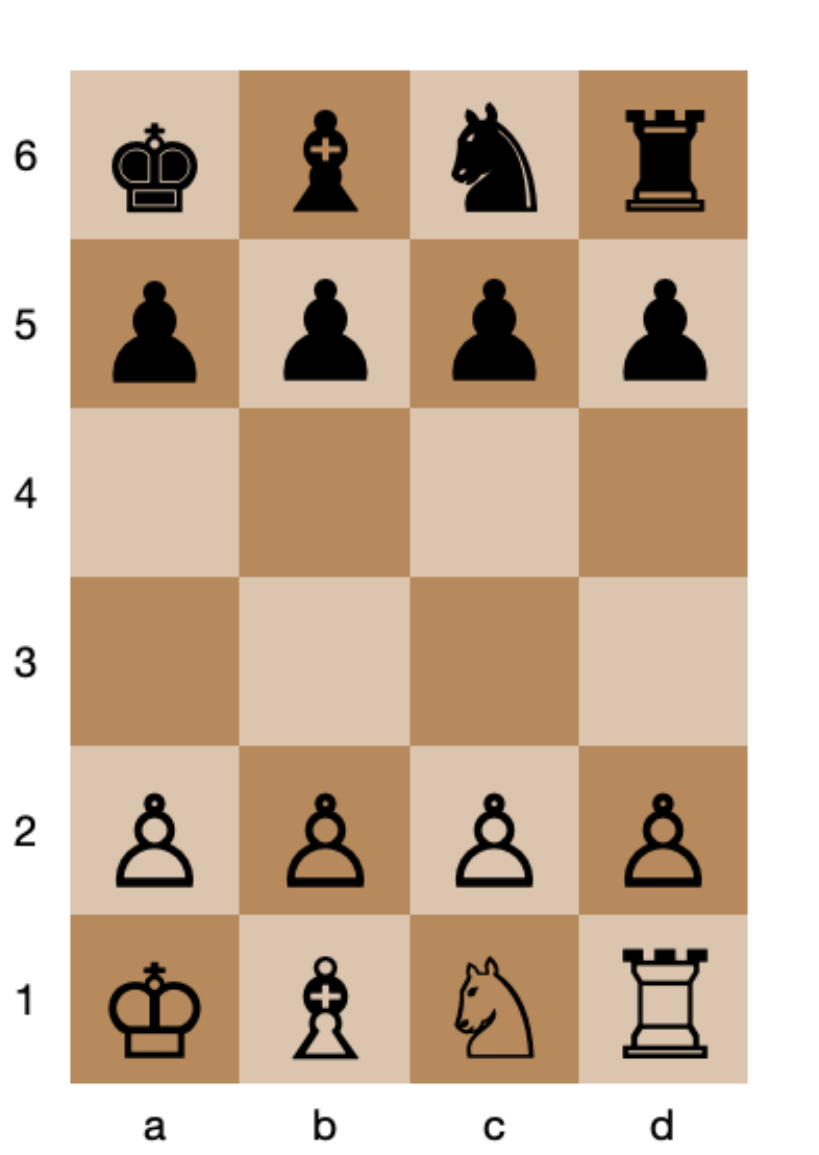}\;\includegraphics[width=0.2776\columnwidth]{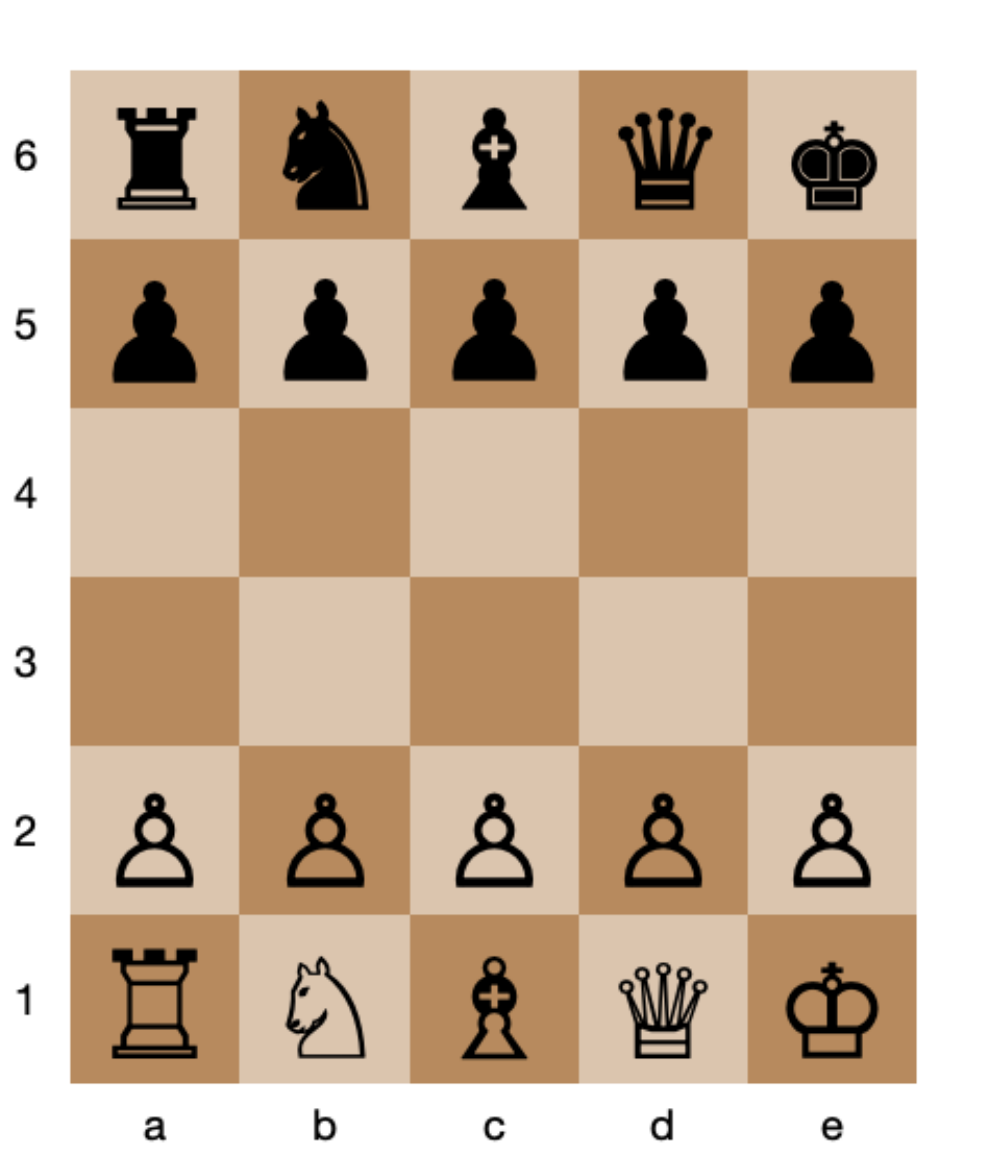}
\caption{Smaller board sizes and initial positions.}\label{figmini}
\end{figure}

Other variants of Niel's Chess can be devised, by directing and interpreting the marks on the indefinite pieces in different ways. The game can also be played on smaller boards~\cite{minichess}, such as on $4\times 4$, $4\times 5$, $4\times 6$ or $5\times 6$ boards with initial positions shown in Figure~\ref{figmini}. From an educational perspective, different quantum phenomena can be illustrated by different rules. For instance, one can exhibit partial collapse by allowing conventional pieces to be on three or more squares at once, or complex amplitudes via fine-grained rotations, or the Pauli-$X$ gate by allowing both pieces of an indefinite pair instance to be rotated by $180$ degrees, or even a spin state with real amplitudes via a single rotated mark. The limit is the imagination.

\begin{figure}[t]
\centering
\includegraphics[width=0.48\columnwidth]{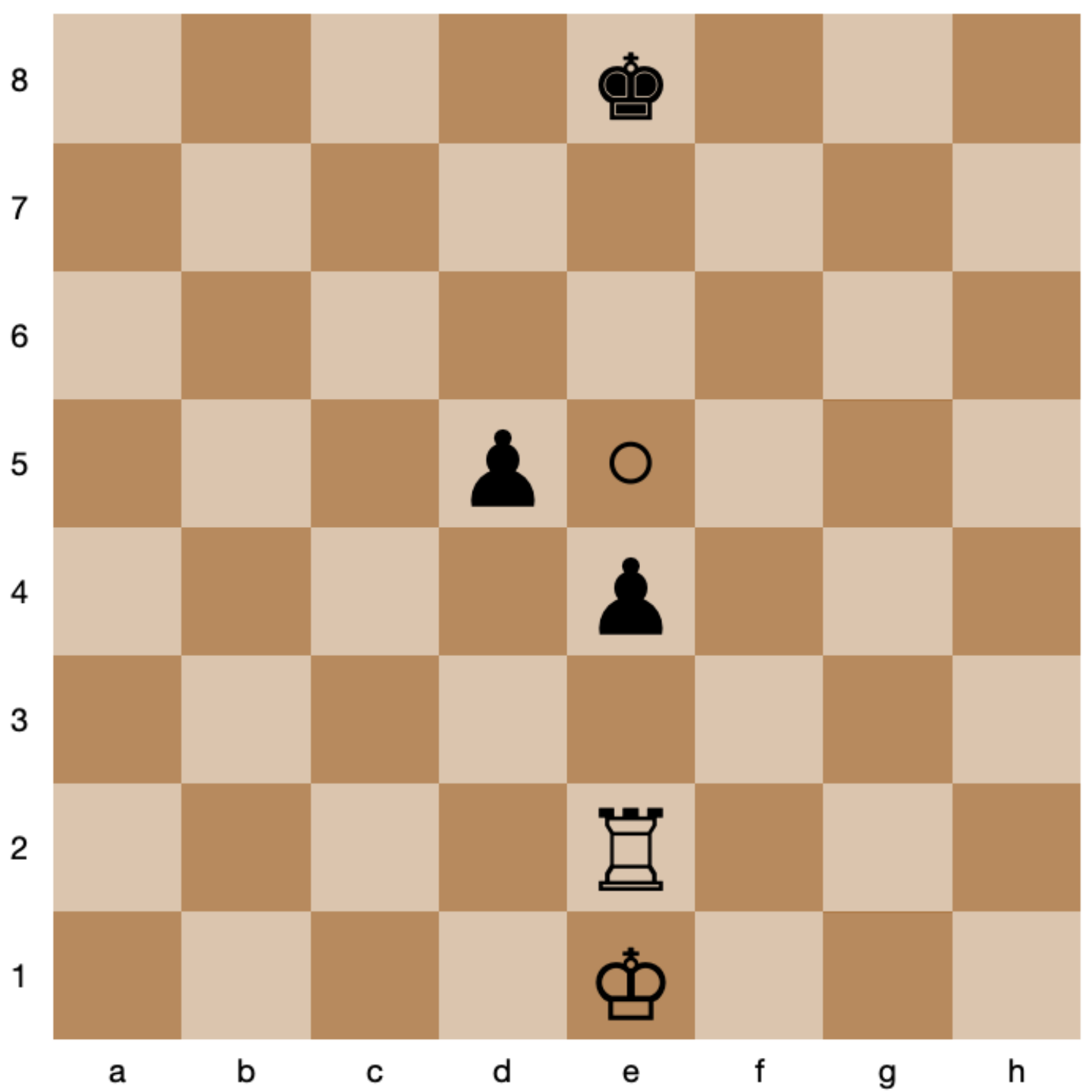}
\caption{The rook attempts to tunnel to e5.}\label{figtunnel1}
\end{figure}

\begin{figure}[t]
\centering
\includegraphics[width=0.48\columnwidth]{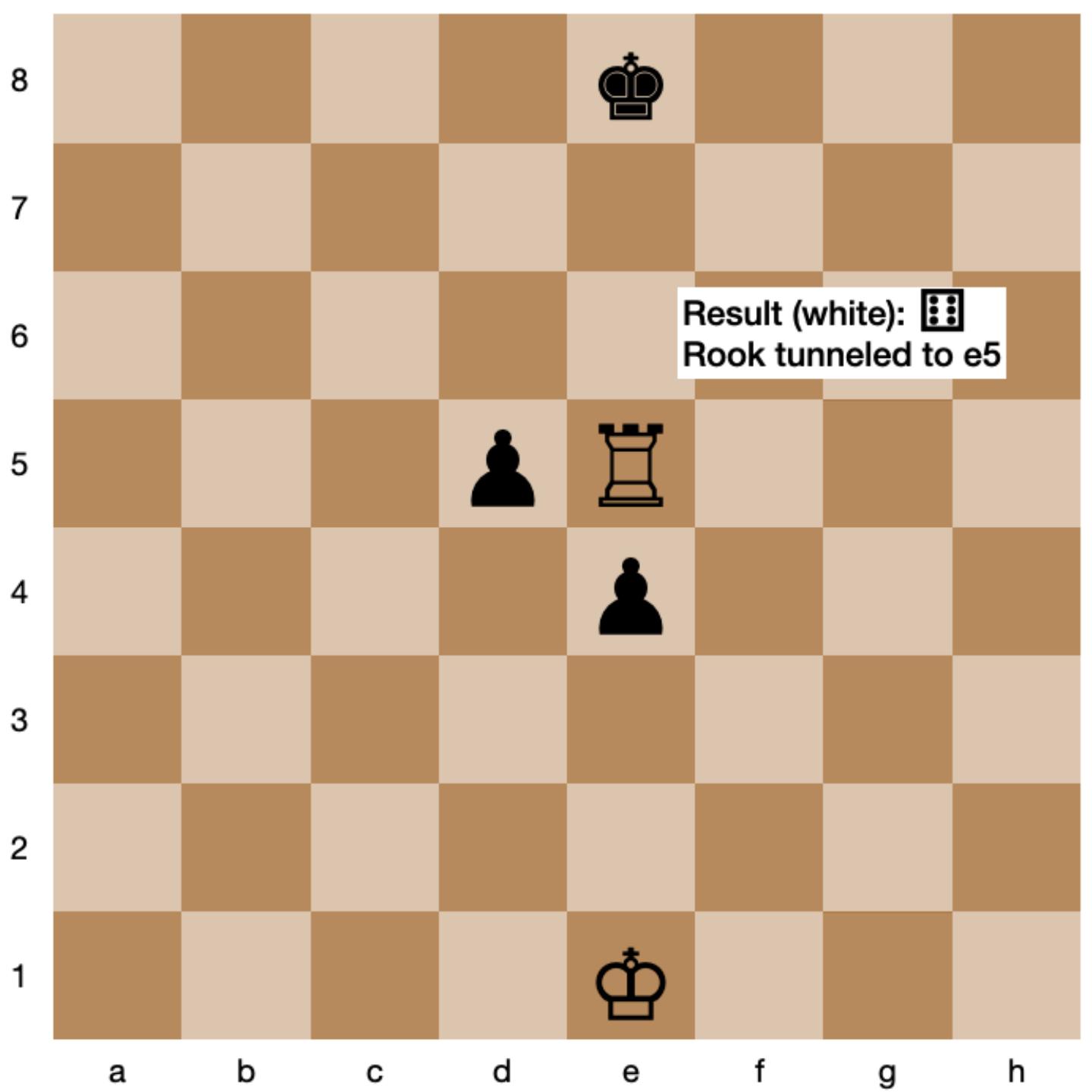}\quad\includegraphics[width=0.48\columnwidth]{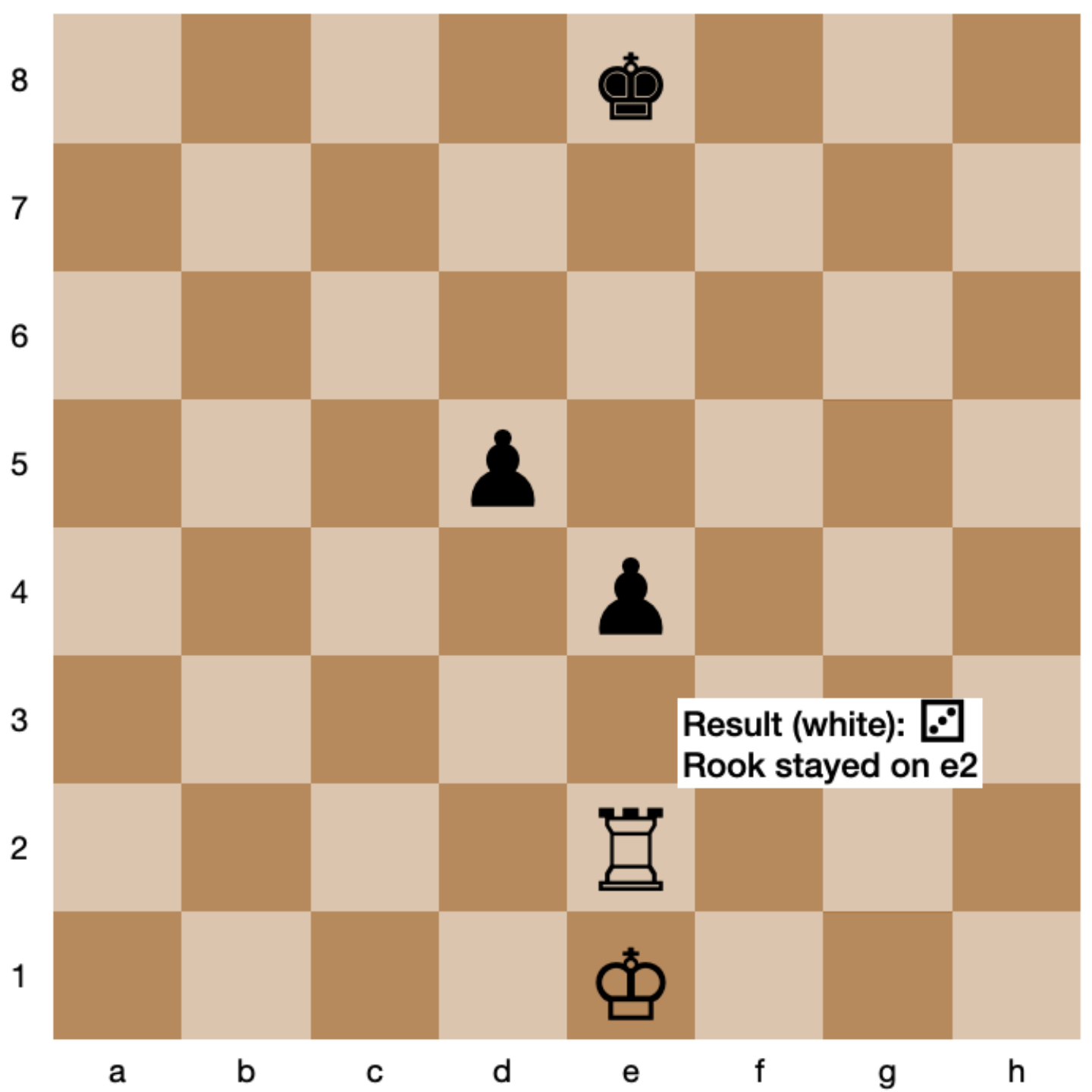}
\centering
\caption{Left: success, White rolled a 6. Right: failed to tunnel, White rolled a 3.}\label{figtunnel2}
\end{figure}

One idea, which was used in the pilot, is quantum tunneling, shown in Figures~\ref{figtunnel1} and~\ref{figtunnel2}. By rolling a 6, the rook successfully tunnels through an opponent's pawn to an unoccupied square right behind that pawn. Otherwise, by rolling 1 to 5 the rook would "bounce back" and stay where it was. Quantum tunneling may be attempted only when both involved pieces are conventional ones of different color, and the targeted square must be right behind the opponent's piece, unoccupied.

\end{appendices}

\end{document}